%% file: main.tex
\documentclass[sigconf]{acmart}
\DeclareUnicodeCharacter{E904}{\textbf{?}}
\AtBeginDocument{%
  }

\setcopyright{acmlicensed}
\copyrightyear{2018}
\acmYear{2018}
\acmDOI{XXXXXXX.XXXXXXX}

\acmConference[Conference acronym 'XX]{Make sure to enter the correct
  conference title from your rights confirmation emai}{June 03--05,
  2018}{Woodstock, NY}
\acmISBN{978-1-4503-XXXX-X/18/06}

\input{codes}

\usepackage[flushleft]{threeparttable}
\usepackage{booktabs}
\usepackage{multirow}
\usepackage[normalem]{ulem} 
\usepackage{nicematrix}
\usepackage{colortbl}
\usepackage{graphicx}

\begin{document}

\title[Human-centered explanation does not fit all]{Human-centered explanation does not fit all: The Interplay of sociotechnical, cognitive, and individual factors in the effect of AI explanations in algorithmic decision-making}

\author{Yongsu Ahn}
\email{yongsu.ahn@pitt.edu}
\affiliation{%
  \institution{University of Pittsburgh}
  \city{Pittsburgh}
  \state{Pennsylvania}
  \country{USA}
}

\author{Yu-Ru Lin}
\email{yurulin@pitt.edu}
\affiliation{%
  \institution{University of Pittsburgh}
  \city{Pittsburgh}
  \state{Pennsylvania}
  \country{USA}
}

\author{Malihe Alikhani}
\email{m.alikhani@northeastern.edu}
\affiliation{%
  \institution{Northeastern University}
  \city{Boston}
  \state{Massachusetts}
  \country{USA}
}

\author{Eunjeong Cheon}
\email{echeon@syr.edu}
\affiliation{%
  \institution{Syracuse University}
  \city{Syracuse}
  \state{New York}
  \country{USA}
}

\renewcommand{\shortauthors}{ Ahn et al.}

\begin{abstract}
  Recent XAI studies have investigated what constitutes a \textit{good} explanation in AI-assisted decision-making. Despite the widely accepted human-friendly properties of explanations, such as contrastive and selective, existing studies have yielded inconsistent findings. To address these gaps, our study focuses on the cognitive dimensions of explanation evaluation, by evaluating six explanations with different contrastive strategies and information selectivity and scrutinizing factors behind their valuation process. Our analysis results find that contrastive explanations are not the most preferable or understandable in general; Rather, different contrastive and selective explanations were appreciated to a different extent based on who they are, when, how, and what to explain -- with different level of cognitive load and engagement and sociotechnical contexts. Given these findings, we call for a nuanced view of explanation strategies, with implications for designing AI interfaces to accommodate individual and contextual differences in AI-assisted decision-making.
\end{abstract}



\received{20 February 2007}
\received[revised]{12 March 2009}
\received[accepted]{5 June 2009}

\maketitle

\input{010_Introduction}
\input{020_Related_work}
\input{030_Overview}
\input{040_Method}

\input{050_Results}
\input{060_Discussion}
\input{070_Conclusion}


\bibliographystyle{ACM-Reference-Format}
\bibliography{references, references-zotero}

\appendix
\input{080_appendix}

\end{document}

%% file: codes.tex
\usepackage[backgroundcolor=white,textsize=tiny]{todonotes}
 
\usepackage{aliascnt}
\usepackage{tabularx, booktabs}
\usepackage{array}

\usepackage{natbib}
\usepackage{multirow}
\usepackage{xspace}
\usepackage{soul}
\usepackage{caption}
\usepackage{placeins}
\usepackage{url}

\newcolumntype{L}[1]{>{\raggedright\let\newline\\\arraybackslash\hspace{0pt}}p{#1}}
\newcolumntype{C}[1]{>{\centering\let\newline\\\arraybackslash\hspace{0pt}}p{#1}}
\newcolumntype{R}[1]{>{\raggedleft\let\newline\\\arraybackslash\hspace{0pt}}p{#1}}

\newcommand{\loanN}{{\sf \small \color{black} Loan-}\xspace}
\newcommand{\drivN}{{\sf \small \color{black} Driv-}\xspace}
\newcommand{\mediP}{{\sf \small \color{black} Medi+}\xspace}
\newcommand{\mediN}{{\sf \small \color{black} Medi-}\xspace}
\newcommand{\recomP}{{\sf \small \color{black} Recom+}\xspace}
\newcommand{\recomN}{{\sf \small \color{black} Recom-}\xspace}

\newcommand{\comp}{{\sf \small \color{black} Complete}\xspace}
\newcommand{\cbhe}{{\sf \small \color{black} Case-based (hetero)}\xspace}
\newcommand{\cbho}{{\sf \small \color{black} Case-based (homo)}\xspace}
\newcommand{\cto}{{\sf \small \color{black} Contrastive (o)}\xspace}
\newcommand{\ctt}{{\sf \small \color{black} Contrastive (t)}\xspace}

\newcommand{\cf}{{\sf \small \color{black} Counterfactual}\xspace}

\newcommand{\darkgray}[1]{{\textcolor[HTML]{8d8d8d}{#1}}}
\newcommand{\orange}[1]{{\textcolor{orange}{#1}}}
\newcommand{\darkorange}[1]{{\textcolor[HTML]{cd5f00}{#1}}}
\newcommand{\green}[1]{{\textcolor[HTML]{679092}{#1}}}
\newcommand{\darkgreen}[1]{{\textcolor[HTML]{006160}{#1}}}
\newcommand{\darkpurple}[1]{{\textcolor[HTML]{4b39ce}{#1}}}

\newcommand{\appsec}[1]{{\textcolor{blue}{{\it Appendix}~}(Section~\ref{#1})}\xspace}

%% file: 010_Introduction.tex
\section{Introduction}
\label{sec:intro}

Artificial intelligence (AI) has increasingly become integrated into everyday life, influencing decisions from personalized purchase recommendations to media content suggestions. However, concerns about AI's potential to produce unfair or harmful outcomes have highlighted the need for greater transparency and explainability in AI-driven decision-making \cite{lipton2018mythos}. In response, the field of eXplainable AI (XAI) has developed various explanation strategies aimed at making AI systems more transparent; However, a critical question still remains: what constitutes a {\it good} explanation in the context of AI decision-making?

Drawing from cognitive and social sciences, Miller \cite{miller2019explanation} introduced human-centered explanation properties---such as selective, contrastive, and conversational explanations---that have since gained considerable attention. Of particular interest are {\it contrastive} explanations, which highlight the differences between a decision and its alternatives, and {\it selective explanations}, which focus on the most relevant information. These strategies have been widely studied, from algorithmic methods to user-facing empirical studies.

\begin{figure}[t]%
\centering
\includegraphics[width=\columnwidth]{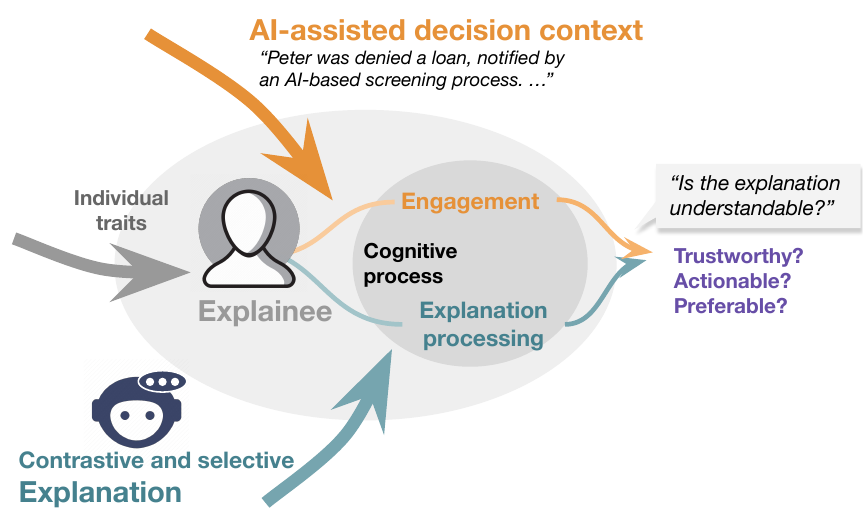}
\caption{The process of humans' valuation of explanations. This study explores the interplay of \darkgray{individual-}, \orange{context-}, and \darkgreen{explanation-dependent} factors influencing the \darkpurple{valuation} of explanation strategies.}
\label{fig:teaser}
\end{figure}

However, despite their theoretical appeal, findings on these explanation types have been inconsistent. For instance, while contrastive explanations have been shown to improve decision quality in certain tasks like emotion recognition \cite{RelatableExplainableAIPerceptualProcess}, they have also failed to align AI trust with human perception in other contexts \cite{AreExplanationsHelpfulComparative}. Similarly, while detailed explanations can enhance users' understanding of AI processes \cite{TooMuchTooLittlea, ExplainableAutonomyStudyExplanation}, they also risk fostering over-reliance on AI decisions \cite{RoleExplanationsTrustReliance} and reducing trust in AI systems \cite{HowMuchInformationEffects}. These mixed findings raise the question: are contrastive and selective explanations universally beneficial, or do their advantages depend on specific contexts?

We argue that the inconsistencies in previous studies stem from insufficient attention to the cognitive and contextual factors that influence how users evaluate explanations. Research in cognitive science suggests that individual traits such as cognitive load and decision-making style play a critical role in how explanations are perceived and valued \cite{PartneringPeopleDeepLearning, SpanishvalidationGeneralDecisionMakingStyle}. Recent XAI literature has touched on the influence of sociotechnical contexts and user diversity on AI transparency \cite{CapturingTrendsApplicationsIssues, ScienceHumanAIDecisionMakingSurvey, ehsan2023charting}, but most studies are limited to single contexts or small sample sizes, preventing a comprehensive understanding of how these factors shape the impact of explanations.

To address these gaps, our study focuses on the cognitive dimensions of explanation evaluation. Specifically, we examine the following:
(1) We evaluate the effectiveness of widely accepted explanation strategies---contrastive and selective---through an experiment involving laypeople. By comparing six explanation styles drawn from existing literature, we assess how different approaches to contrast and selectivity influence user preferences.
(2) We explore how individual, contextual, and cognitive factors shape the evaluation process. Drawing on insights from cognitive and social science, we synthesize these factors into a comprehensive framework to explain when, why, and for whom specific explanation strategies are most effective.

\textbf{Study contexts.} While a variety of contexts in AI-assisted decision-making has been studied in existing literature, we particularly focus on the context of algorithmic decision-making that determines various aspects of individuals’ everyday lives, such as loan approval, medical diagnosis, autonomous driving, and movie recommendations. These contexts typically entail a passive nature of decisions made by algorithms embedded within AI systems or infrastructures in various socio-technical contexts, where decisions are notified or presented to data subjects, triggering their cognitive responses or emotions on the given decisions. This contrasts with the general or more objective nature of AI-assisted decision-making contexts, such as image/sentiment classification, where explanations can assist with better understanding of feature attribution. Due to this nature, the design of our study, which best approximates the context with written scenarios then asks subjective ratings of users, centers on exploring their perceived and cognitive responses, along with the various factors that shape them, particularly when explanations describe them, contain personal information, and contrast with others’.

Overall, our findings challenge the assumption that selective and contrastive explanations are universally intuitive or advantageous. Instead, we demonstrate that their effectiveness is highly dependent on context and user characteristics. This evidence supports the need for a more context-sensitive, user-centered approach to designing XAI systems. The key contributions of our study are as follows:
\begin{itemize}
\item {\bf Context-specific values of explanations:} Our study reveals that the perceived value of contrastive and selective explanations varies significantly by context and individual traits, challenging the assumption that these strategies are inherently human-friendly.
\item {\bf Novel framework for explanation evaluation:} We introduce a comprehensive framework (Fig. \ref{fig:teaser}) that captures the complex interplay of \darkgray{individual differences}, \orange{sociotechnical contexts}, and \darkgreen{cognitive aspects} in explanation valuation. While our study focuses on specific types of explanations, this framework can be applied broadly across various AI decision-making contexts.
\item {\bf Design implications for XAI:} We offer actionable insights for designing more interactive and personalized XAI systems. Our recommendations emphasize the need to adapt explanation strategies based on user characteristics and contextual factors, improving both the usability and trustworthiness of AI systems.
\end{itemize}

%% file: 020_Related_work.tex
\section{Related Work}
\label{sec:related-work}

\subsection{Valuation of Explanation Strategies in AI-assisted Decision-making}
\label{sec:related-work-exp}

Evaluating post-hoc explanations in the form of text has gained much attention as it provides rationales on AI-based decisions and impacts users' perceived trust and understandability  \cite{vilone2020explainable, lipton2018mythos, ExplainableSoftwareAnalytics}. Such explanations incorporating a variety of strategies and logics vary by techniques \cite{AreExplanationsHelpfulComparative, adhikari2019leafage} including feature importance or nearest neighbors or logics/strategies such as contrastive \cite{miller2019explanation, ConversationalProcessesCausalExplanation, DeductiveApproachCausalInference}, counterfactual \cite{Contrastscounterfactualscauses, CounterfactualStoryReasoningGeneration, CounterfactualExplanationsOpeningBlack, PsychologicalStudiesCausalCounterfactualReasoning} explanations in social and cognitive science, and case-based explanations from expert system studies \cite{Similaritymeasuresattributeselectioncasebased, GainingInsightCasebasedExplanation, EvaluationUsefulnessCaseBasedExplanation, SurveyCBRApplicationAreas}. 

However, recent XAI research has produced contradictory findings regarding the effect of explanation strategies. For example, providing such explanations in AI-assisted decision were found to help justify the decision and increase users' trust when compared with no explanation provided in some contexts such as medical chatbot \cite{ExploringPromotingDiagnosticTransparency} or self-driving context \cite{TrustingXAIEffectsdifferenttypes}. On the other hand, they tend to be distracting or lead to either overreliance \cite{ExplanationsCanReduceOverrelianceAI, PartneringPeopleDeepLearning} or cognitive overload \cite{WaitWhyAssessingBehaviorExplanation, EvolutionCognitiveLoadTheoryMeasurement} for lay users without a careful design of how explanations are presented \cite{CheXplainEnablingPhysiciansExploreUnderstand}. 
The impact of different explanation strategies in AI-assisted decision-making is also diverging. For example, complete explanations with a greater information complexity, for example, were proved to gain trust better in medical diagnosis \cite{EffectExplanationStylesUser, TooMuchTooLittlea, HowMuchInformationEffects} while they led to over-reliance on AI in medical diagnosis in \cite{RoleExplanationsTrustReliance}. Contrastive explanations known as intuitive and human-friendly were helpful in improving decision quality with semantic evidence in emotion recognition task \cite{RelatableExplainableAIPerceptualProcess}, while it was found to fail to calibrate the trust of AI in human perception in \cite{AreExplanationsHelpfulComparative}.

A recent XAI study \cite{SelectiveExplanationsLeveragingHumanInput} has advanced the discussion on making explanations more selective. The proposed framework draws on literature exploring how humans produce explanations selectively, focusing on aspects such as abnormality \cite{hilton1986knowledge}, relevance \cite{woodward2006sensitive}, and changeability \cite{hilton2007course}. Empirical results from this study suggest that allowing users to contribute to the generation of explanations reduces over-reliance on automated systems.

In this study, we pursue a deeper and comprehensive understanding of users' valuation on explanations strategies. By conducting a survey-based experiment, we find that the valuation process is a complex interplay of individual and contextual factors, highlighting the importance of personalizing the degree of explainability carefully based on individual traits and cognitive abilities.

\subsection{Individual and Cognitive Dimension of Explanatory Process}
\label{sec:related-work-cognition}

When making sense of a social situation such as interpersonal communication or decision-making processes, individuals go through certain cognitive processes to analyze, interpret, and remember information \cite{SocialCognition}. Previous studies found that these process are influenced by individuals’ own cognitive tendency or given contexts. For example, people tend to have their own decision-making style \cite{DecisionmakingstylesreallifedecisionChoosing,Individualdifferencesadultdecisionmakingcompetence, DecisionMakingStyleDevelopmentAssessmentNew} -- whether they rely on hunches or a thorough search for information. According to the Motivation-Opportunity-Ability (MOA) model \cite{EnhancingMeasuringConsumersMotivationOpportunity}, individuals are either empowered or hindered in exhibiting behavioral changes or attributing decisions when making sense of their success and failure in their careers or education \cite{weiner1972attribution}. These cognitive processes can further influence the degree to which individuals seek more information to reason about situations, manifest in either spontaneous or deliberate modes within the dual processing theory \cite{SocialCognition, CogitoergoquidEffectCognitive, InfluenceCognitiveStylesUsersUnderstandinga}.

A number of studies \cite{PartneringPeopleDeepLearning, Doeshighereducationhonecognitive, Domainspecificpreferencesintuitiondeliberationdecision} have found that all these cognitive traits are highly dependent on their demographics. For example, older adults rely more on emotions and experience rather than being rational due to aging cognitive ability \cite{lockenhoff2018aging}. The level of education also influences the cognitive load, intelligence, thinking, and working memory \cite{Doeshighereducationhonecognitive}.  Rational thinkers tend to actively seek explanations when finding the best recommendations than intuitive thinkers. 

Despite these findings, there is limited understanding of the complex cognitive dimensions that affect individuals' willingness to seek explanations and their valuation of specific explanation strategies in AI-assisted decision-making. Our study explores these connections, highlighting the need for careful designed explanations that consider these cognitive processes.

\subsection{Evaluating Explanations in Various AI-assisted Decision Contexts}
\label{sec:related-work-context}

The effectiveness of various explanation types in enhancing trust and understanding has been studied across diverse AI-assisted decision contexts, including human-robot interaction \cite{ExplainableAgentsRobotsResults, GuidelinesDevelopingExplainableCognitive, SelfExplainingSocialRobotsVerbal, TheoryExplanationsHumanRobotCollaboration, DifferentXAIDifferentHRI}, self-driving technologies \cite{ExplainableAutonomyStudyExplanation, TextualExplanationsSelfDrivingVehicles, TrustingXAIEffectsdifferenttypes}, and medical diagnosis \cite{humanbodyblackboxsupporting, EffectExplanationStylesUser, jimenez2020drug}.

For instance, in medical context, providing explanations helped improve health awareness, facilitate learning, and aid decision-making by offering patients new information about their symptoms \cite{ExploringPromotingDiagnosticTransparency} or help laypeople understand complex medical concepts during cancer diagnosis \cite{EffectExplanationStylesUser}. In self-driving contexts, explanations provided in a timely manner during sequential driving scenes have been found to improve understanding \cite{TrustingXAIEffectsdifferenttypes, ExplainableAutonomyStudyExplanation, TextualExplanationsSelfDrivingVehicles}.

Despite all these studies highlighting the impact of various explanation styles within a certain context, it remains uncertain how different explanation types affect the levels of trust, understandability, and other aspects of user perceived values across multiple contexts. While previous studies \cite{CapturingTrendsApplicationsIssues, ScienceHumanAIDecisionMakingSurvey, ehsan2023charting} have conceptually examined that AI transparency is entangled with sociotechnical contexts, our research empirically demonstrates that the effectiveness of explanations varies based on the application context, highlighting the need for context-aware design in explainable AI systems.

%% file: 030_Overview.tex
\section{Study Purpose and Overview}
Our study aims to examine two key aspects of explanation strategies in AI-assisted decision-making: (1) how contrastive and selective explanations—commonly accepted as properties of good explanations—are perceived by laypeople, and (2) which factors influence the evaluation of these explanation strategies in AI-assisted decisions. We explore various AI-assisted decision-making scenarios, where human subjects receive decisions related to their personal circumstances in different sociotechnical contexts. These decisions are accompanied by textual explanations that provide the rationale behind the outcomes.

To ground our study in concrete examples, we developed a range of written scenarios across various decision-making contexts, such as medical diagnoses and loan approvals. These scenarios demonstrate how explanations are provided and how participants might evaluate them in different sociotechnical contexts. Below, we present one of the six selected scenarios.



\textbf{Loan approval case as a motivating scenario:} On a Tuesday morning, Jane, a sixty-seven-year-old retiree, applied for a loan to renovate her home. A few hours later, she received a notification from the bank stating that her loan request had been denied. The AI-based system explained the decision was based on her credit score and financial history. This result surprised Jane, as she had always been responsible with her finances. She wanted to understand why the loan was denied---whether it was her age, retirement status, or something else in her profile.

\input{tables/tab_exp_strategies}

Jane found an AI-based chatbot on the bank's website offering explanations for loan decisions. She asked the chatbot why her loan was denied, and it provided a detailed response outlining the factors behind the decision.

\subsection{Evaluation of Contrastive and Selective Explanations}

In such scenarios, AI systems must not only provide explanations to meet explainability requirements but also tailor them to users' varying characteristics, levels of understanding, and expectations within AI-assisted decision-making contexts.

Our selection of explanation types, informed by a thorough literature review, enables us to examine different forms of contrastive and selective explanations. As summarized in Table \ref{tab:strategies}, these explanations fall into two main categories: contrastive strategies and information selectivity.

{\bf Contrastive strategies.} Contrastive explanations compare a given decision to alternative outcomes using different strategies ({\it how} to compare). These strategies include comparing to an opposite outcome (contrastive), comparing to a similar case (analogous), or providing no comparison at all. The focus of the comparison ({\it what/who} to compare) can also vary, such as comparing the user's outcome to someone else's (\cto) or comparing their current status to their past (\ctt) \cite{miller2019explanation, RemoteCausesBadExplanations}, or considering a hypothetical alternative outcome they could face (\cf). Each of these variants is rooted in the existing literature (see detailed review in Section \ref{sec:related-work-exp}) and represents a specific approach to helping users understand the AI's decision-making process.

{\bf Information selectivity.} The explanations in our study are also differentiated by information selectivity, which involves two sub-components: information complexity and alignment with users' prior beliefs. 

\begin{figure*}[h]%
\centering
\includegraphics[width=\textwidth]{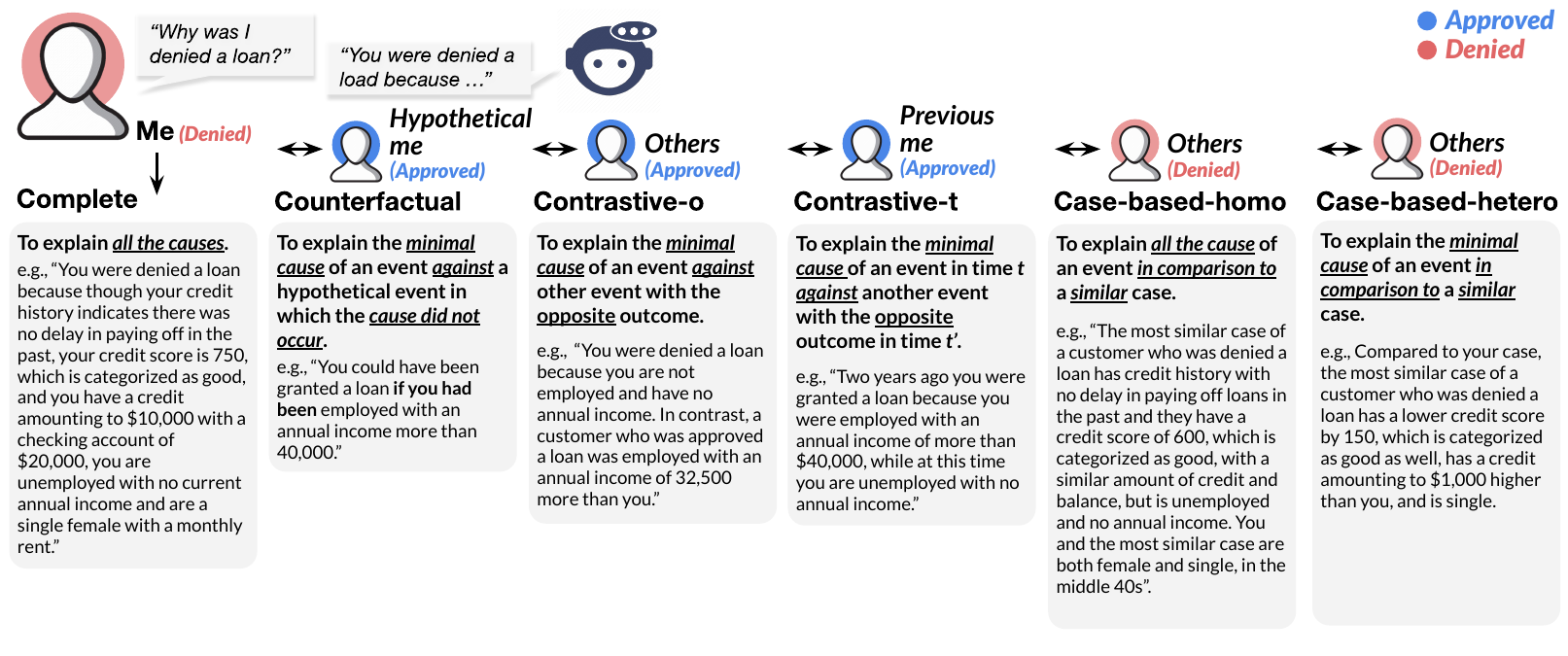}
\vspace{-2em}
\caption{The detailed definitions and examples of six variants of explanations.}\label{fig:exp-strategies}
\end{figure*}

In terms of information complexity, some explanations, such as contrastive and counterfactual explanations (\cto, \ctt, and \cf) and \cbhe \cite{ EvaluationUsefulnessCaseBasedExplanationa, stolpmann1999optimierung}, provide simple explanations that highlight only the minimal features differentiating the cases. In contrast, other explanations, like \comp and \cbho \cite{el2015case, ExplanationsCaseBasedReasoningFoundational}, offer a comprehensive overview, describing all relevant features in detail. This variation further influences how much and what information is going to be described in each explanation. Figure \ref{fig:exp-strategies} illustrates six different explanation strategies in the context of a loan denial decision. As shown, explanations with minimal information focus on a few key features, while thorough explanations include all relevant factors.

We hypothesize that the perceived value of these explanations depends on how well the selected features align with users' prior beliefs. This relationship can be compared to the trade-off between bias and variance in prediction. Presenting minimal information may result in higher bias (less alignment with users' expectations), but when the explanation aligns with their expectations, it can lead to greater satisfaction due to its greater precision. In contrast, presenting more comprehensive explanations tends to reduce bias but may introduce greater variance, potentially overwhelming users with information.

\subsection{Understanding of Human Valuation Process of Explanations} 
Given an array of contrastive and selective explanations, which explanation strategies will people find preferable and intuitive? While theoretical frameworks advocate for these types of explanations as generally intuitive and human-friendly, we posit that their value is highly contextual — depending on factors such as \textit{when}, \textit{how}, and \textit{for whom} the explanations are provided.

\begin{figure*}[h]%
\centering
\includegraphics[width=\textwidth]{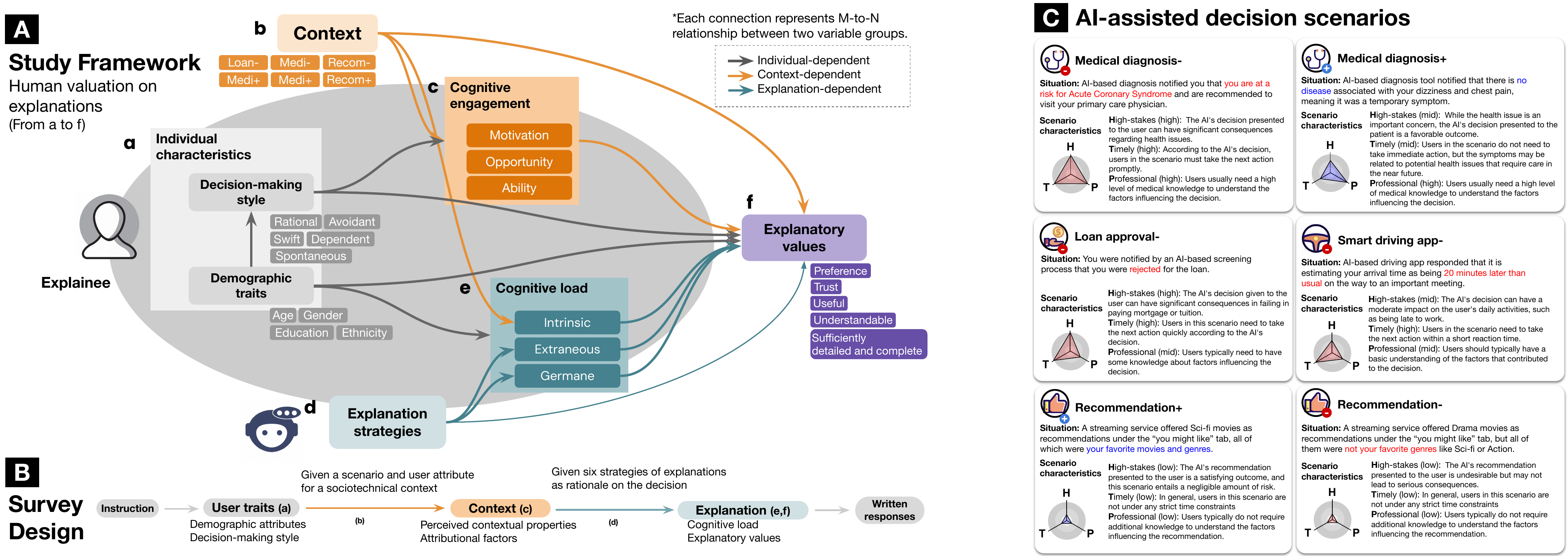}
\caption{A) \textbf{Study framework}: Our study aims to systematically explore the inner workings of how individuals attribute AI-generated decisions when confronted with AI-driven decisions (e.g., why was I denied a loan?) and evaluate explanations presented by AI systems. The study framework for human valuation on explanations presents a cognitive journey (in the order from a to f) explainees may go through while processing explanations about the decision. B) \textbf{Survey design}: To examine the valuation process, we distilled the framework into a survey for scenario-based experiment. C) \textbf{AI-assisted decision scenarios}: These decision scenarios are carefully selected to be contrasted in three aspects: high-stakes, professional, and timely.}
\label{fig:study-design}
\end{figure*}

Prior research in cognitive and social science has presented numerous findings on how individuals engage with explanatory processes or decision-making contexts differently based on their personal traits and situational contexts. Drawing from the literature review (see details in Section \ref{sec:variable-selection}), we present the framework for conceptualizing human valuation of explanation. As presented in Fig. \ref{fig:study-design}A, we synthesize them as an interplay of various factors, which can fall into one of the following dimensions: 1) \darkgray{individual-dependent} factors: Individuals are known to have inherent \darkgray{demographic traits} and \darkgray{decision-making style} (Fig. \ref{fig:study-design}A-a) that shape their preferences and ability based on their inherent traits such as demographics and decision-making style. 2) \orange{context-dependent} factors: When given a decision in a \orange{sociotechnical context} (e.g., denied a loan) (Fig. \ref{fig:study-design}A-b), individuals may perceive the context (e.g., does the decision have a significant consequences?), in turn exhibiting a different level of \darkorange{cognitive engagement} (Fig. \ref{fig:study-design}A-c), depending on whether they are motivated, perceive a certain level of opportunities or ability to seek explanations. When presented with an \green{explanation} from an AI system (Fig. \ref{fig:study-design}A-d), individuals may experience different levels of \darkgreen{cognitive load} (Fig. \ref{fig:study-design}A-e), impacting their interpretation of the provided explanation. Depending on these factors, users will evaluate explanations regarding different \darkpurple{explanatory values} (Fig. \ref{fig:study-design}A-f). In our study, we scrutinize this framework by investigating RQ3 and RQ4 as outlined in Section \ref{sec:rqs}, aiming to identify key aspects of the human valuation process of explanations.

\subsection{Research Questions}\label{sec:rqs}
Based on two facets of our study purposes as described above, we distill them into four key research questions listed below: \\

{\bf Evaluating laypeople's preferences for contrastive and selective explanations:} The first part of our study focuses on evaluating preferences for contrastive and selective explanations across various AI-assisted decision-making scenarios. We use both quantitative and qualitative approaches: a survey experiment to measure preferences and an analysis of responses to open-ended questions to explore the reasons behind these preferences (i.e., what features of the explanations influenced their choices) (see detailed methods in Sections \ref{sec:study-design} and \ref{sec:method-open})). Through this, we aim to critically examine the commonly held assumption that contrastive and selective explanations are inherently valuable in all contexts.
\begin{itemize}
    \item \textbf{RQ1.} How are contrastive and selective explanations preferred in general and within specific sociotechnical contexts?
    \item \textbf{RQ2.} Which aspects and properties of explanations are linked to the distinct valuation of those explanation strategies?
\end{itemize}

{\bf Exploring factors influencing preferences for contrastive and selective explanations:} In addition to evaluating preferences, we aim to gain deeper insights into the factors that shape how people value different explanations. In our experiment, we translate these factors into specific survey items and analyze their interactions, examining both individual variables and their combined effects. This design seeks to explain the factors influencing preferences for contrastive and selective explanations.

\begin{itemize}
    \item \textbf{RQ3.} How do individual, contextual, and cognitive factors interact in the valuation process of explanations?
    \item \textbf{RQ4.} How do these factors collectively influence the preference on valuation of contrastive and selective explanations?
\end{itemize}

\subsection{Theoretical Foundation and Rationale behind Variable Selection} \label{sec:variable-selection}
To design the framework, we have reviewed literature throughout a wide range of disciplines. Studies in social and cognitive science have been to examine theories and findings about how people attribute decisions and process information based on a given context, leading to shaping individual, cognitive, and contextual factors. An extensive review from and outside of XAI literature have led us to examine explanatory values (i.e., criteria for evaluating explanations in AI contexts).

\textbf{From various disciplines outside of XAI literature.} Literature from social and cognitive science have established theories and findings on individuals' cognitive capability and status in the decision-making and explanatory process. Our objective is to integrate these insights into AI decision contexts, to not only theoretically support our framework but also make connections between human and AI studies, from human-human and human-AI communication.

\begin{itemize}
    \item \textbf{\darkgray{Decision-making styles}} (Fig \ref{fig:study-design}A-a): Research in Psychology and Business has studied how individuals have different levels of cognitive tendency based on inner traits such as gender, age, or education. Especially when processing decisions and relevant information, individuals tend to exhibit different decision-making styles (See detailed literature review in Section \ref{sec:related-work-cognition}). We introduce five different styles in a scale called General Decision Making Style (GDMS) \cite{examinationgeneraldecisionmakingstyle}, including rational, avoidant, intuitive, dependent, and swift (details in Table \ref{tab:cog_vars}).
    \item \textbf{\darkorange{Cognitive engagement}} (Fig \ref{fig:study-design}A-c): In the literature from social science such as Education and Consumer Behavior, scholars have attempted to establish theories, often referred to as social attribution, theorizing individuals' internal cognitive states that drive behavioral changes (See detailed literature review in Section \ref{sec:related-work-cognition}) triggered by events such as a failure or success of their career or educational performances. In this study, we employ the MOA (Motivation, Opportunity, Ability) model \cite{EnhancingMeasuringConsumersMotivationOpportunity, jepson2018applying, fazio2014attitude} integrating three internal states, indicating whether they are more willing to act or change when they feel more motivated, perceive more opportunities or ability depending on surrounding environments or situations (details in Table \ref{tab:cog_vars}).
    \item \textbf{\darkgreen{Cognitive load}} (Fig \ref{fig:study-design}A-e): Cognitive load theory is a widely used construct to measure cognitive burden in humans’ information processing  (See detailed literature review in Section \ref{sec:related-work-cognition}). We examine three facets of cognitive load, including intrinsic, extraneous, and germane load. Each facet of cognitive load serves a distinct purpose: Intrinsic load is contingent on the difficulty of information covered in explanations, heavily influenced by specific contexts. On the other hand, extraneous and germane load pertain more to the presentation of information, and this variability is tied to the explanation strategies (details in Table \ref{tab:cog_vars}).
\end{itemize}

\textbf{From XAI literature.} Recent XAI studies have examined AI explainability specific to a context or multiple contexts to evaluate its effect in enhancing various explanatory values.

\begin{itemize}
    \item \textbf{\darkpurple{Explanatory values}} (Fig \ref{fig:study-design}A-f): We examined XAI surveys \cite{MetricsExplainableAIChallenges, TaxonomyHumanSubjectEvaluation, ScienceHumanAIDecisionMakingSurvey} investigating measures, criteria, and taxonomy for evaluating explainable AI. In our study, we employ multiple values for users to evaluate the effect of explanations, whether explanations have sufficiency (detailed and complete), understandability (helping understand why the decision was made), usefulness (facilitating comprehension of actions to take), trust (increasing the willingness to act on the basis of, the recommendations, actions, and decisions of an artificially intelligent decision aid)\footnote{A recent study \cite{HowEvaluateTrustAIAssistedDecision} provides the definition of trust in evaluating XAI systems and how it should be integrated in the experimental protocols. We note that our experimental setting and measure satisfy the three elements of trust discussed in the paper, including attitude (i.e., measures in the experiment capture how participants perceive the AI system), vulnerability (i.e., the given scenario involves uncertainty of the outcomes of a decision), positive expectation (i.e., the instruction ensures participants that the AI system is reliable and accurate).}, and overall preference.
    \item \textbf{\orange{Sociotechnical contexts}} (Fig \ref{fig:study-design}A-b): The six decision scenarios (Fig. \ref{fig:study-design}C) were selected to examine the impact of sociotechnical contexts. These decision contexts were carefully selected to differentiate between them in three contextual aspects: (a) {\it high-stakes}: a decision in the scenario involves significant consequences, (b) {\it professional}: a decision in the scenario requires one to have the professional knowledge to reason the given information, and (c) {\it timely} (or {\it time-critical}): a decision in the scenario must be made promptly.

    \indent We selected four application contexts widely studied in AI explanability studies including loan approval, medical diagnosis, driving app, and movie recommendation. We further differentiate them with the outcome of a decision -- whether the decision is favorable (positive) or unfavorable (negative) -- may alter people's expectations regarding the explanations. As illustrated in Fig.~\ref{fig:study-design}C, we create six scenarios featuring the various aspects: loan decision (\loanN) and driving app (\drivN) with undesirable decision, medical diagnosis with desirable (\mediP) and undesirable decision (\mediN), desirable (\recomP) and undesirable (\recomN) movie recommendation. The positive and negative signs indicate whether the provided information is generally desirable or undesirable.
\end{itemize}

%% file: tables/tab_exp_strategies.tex
\begin{table*}[h]
\caption{Types of explanation strategies.}
\label{tab:strategies}
\setlength{\tabcolsep}{3pt}
\centering
\begin{tabular}{@{}lllll@{}}
\toprule
\multirow{2}{*}{Explanations} & \multicolumn{2}{l}{Contrastive strategies}                                                                                                                         & \multicolumn{2}{l}{Information selectivity}                                                                                                            \\ \cmidrule(l){2-5} 
                              & \begin{tabular}[c]{@{}l@{}}How to compare\end{tabular} & What/Who to compare                                                                                     & \begin{tabular}[c]{@{}l@{}}Information complexity\end{tabular} & \begin{tabular}[c]{@{}l@{}}Information alignment with \\  prior beliefs\end{tabular} \\ \midrule
\comp          & No strategy                                              & -                                                                                                       & Thorough (All the causes)                                        & Low bias, High variance                                                             \\
\cf            & Contrastive                                              & vs. me (hypothetical status of myself)                                                                  & Simple (Only minimal causes)                                     & High bias, Low variance                                                             \\
\cto           & Contrastive                                              & vs. others (with the opposite outcome)                                                                  & Simple (Only minimal causes)                                     & High bias, Low variance                                                             \\
\begin{tabular}[c]{@{}l@{}}\ctt \\ \newline  \end{tabular}        & \begin{tabular}[c]{@{}l@{}}Contrastive   \\ \newline   \end{tabular}                                        & \begin{tabular}[c]{@{}l@{}}vs. me (previous status of myself \\ with the opposite outcome)\end{tabular} & \begin{tabular}[c]{@{}l@{}}Simple (Only minimal causes)  \\ \newline \end{tabular}                                   & \begin{tabular}[c]{@{}l@{}} High bias, Low variance  \\ \newline  \end{tabular}                                                         \\
\cbhe          & Analogous                                                & vs. others (with the same outcome)                                                                      & Simple (Only minimal causes)                                     & High bias, Low variance                                                             \\
\cbho          & Analogous                                                & vs. others (with the same outcome)                                                                      & Thorough (All the causes)                                        & Low bias, High variance                                                             \\ \bottomrule
\end{tabular}
\begin{tablenotes}
  \small
  \item *Note that explanation strategies that entail contrasting the information (\cf, \cto, and \ctt) inherently deal with simple information by their definition.
\end{tablenotes}
\end{table*}

%% file: 040_Method.tex
\section{Method}\label{sec:method}

\subsection{Study Design}\label{sec:study-design}

We designed a survey experiment to investigate the process of evaluating explanation and answer the research questions. The survey was designed as a between-subject experiment, where participants were randomly assigned to one of the six decision scenarios.

Overall, the survey was designed to emulate the process of humans' valuation of explanation strategies (Fig. \ref{fig:study-design}A), consisting of multiple stages (a-f) as illustrated in Fig. \ref{fig:study-design}B (see the details for the survey material in the \appsec{sec:survey}): a) Participants were asked questions about their individual characteristics including demographics and decision-making style; b) A randomly assigned scenario was presented to engage participants in the AI-assisted decision-making context. To devise each scenario, we consulted with domain experts to determine the details of the outcomes and features, and various user cases that serve as counterparts of the focal data subject regarding contrastive and analogous comparisons; c) Questions regarding perceived contextual properties (i.e., high-stakes, professional, timely) and cognitive engagement (i.e., motivation, opportunity, ability) were prompted to examine context-dependent factors; d) Six explanation strategies were presented; e,f) users were asked to rank explanations to rate their cognitive load as well as explanatory values. Finally, we collected participants' opinions and reasoning for the explanation variants in free-form text responses.


\subsection{Study Implementation}
\label{sec:study-procedure}

\paragraph{Participants}\label{sec:participants}
We recruited participants from Prolific crowdsourcing platform. Workers living in the US (age 18+) are fluent in English were eligible to participate in this study to make sure their ability to reason about the given scenario in English. One of the three rounds aimed to recruit senior people (age 55 and older), as they are typically underrepresented on the crowdsourcing platform. A total of 839 participants took part in the study. We excluded respondents who left their responses as default or who did not respond to all survey questions. This yielded a final sample of 698 participants for our data analysis (390 females, age: 222 participants were between the ages of 18 and 24 and 62 were 65 or older; 1 participant preferred not to state age, and 3 participants preferred not to say their gender). Fig. \ref{fig:demographics} in the \appsec{sec:demo-breaks} provides the demographic breakdown of the participants. The experiment was approved by the Institutional Review Board of the University\footnote{The university name was omitted for blind review.}. We pre-registered the experiment on the Open Science Framework (OSF)\footnote{\url{https://osf.io/hp86t/?view_only=f49fb230b8e8478288d4844869a88863}}. 

\paragraph{Sample size determination}
We used the Mann-Whitney test to determine the sample size. Based on the power analysis with a significance level of 0.05 and a medium effect size, we determined that at least 106 samples per group were needed, resulting in a suggested minimum sample size of 636 for six survey variants, in order to ensure that the effect of the explanations could be tested.

\paragraph{Procedure}
The survey experiment was conducted on Prolific for three rounds in 2022, on January 29, March 29, and April 25. Participants receive a small amount of compensation with the base rate of \$7.74 per hour in exchange for their effort. The survey consisted of 19 questions, seven of which asked for demographic and survey identification information. The median time for a respondent to complete the survey was 11.6 minutes. 

In the context of AI-based decision-making, participants were tasked with evaluating various types of explanations as decision-making rationales. The survey began with questions regarding demographics. Participants were then asked to read a paragraph describing one of six AI-assisted decision-making scenarios provided by the researchers. Given the context, participants first rated their perceptions of contextual properties and cognitive engagement when presented with the scenario. These questions use a five-point  Likert scale with values ranging from 1 to 5. Participants were then instructed to read each explanation style presented in random order and rank them according to their preferences. At the end of the study, participants provided written comments on the rationale behind their choice of the preferred explanations and thoughts about the AI system.

\subsection{Statistical Analyses}
\label{sec:stat}

We conducted multiple types of statistical analyses to examine the research questions listed in Section \ref{sec:rqs}. As a pre-processing step, we converted all ranking responses to a {\it relative rating} variable, such that higher scores indicate positive cognitive loads and explanatory values. 

For RQ3, we employed the Mann-Whitney U test to determine if significant differences existed between two distinct sets of scenarios (e.g., high-stakes vs. low-stakes contexts). The Wilcoxon signed-rank test was additionally utilized for identifying significant differences between pairs of styles, acknowledging that the relative ratings across these styles are interdependent. In our pairwise testing, we applied the Bonferroni correction to adjust for multiple comparisons. Given that our analysis involved both 6 scenarios and 6 styles, the significance level was accordingly adjusted to 0.05/6, which is approximately 0.0083.

To examine RQ4 in scrutinizing the interplay of individual-, cognitive-, and explanation-dependent factors, we used Structural Equation Modeling (SEM) to take into account all direct and indirect relationships between variables included in our study framework (Fig. \ref{fig:study-design}A). For these models per explanation type, we converted participants' ratings of explanation styles into a binary variable indicating whether a particular style was absolutely favorable (i.e., ranked as top or second one) or not.

\subsection{Analysis for Open-ended Questions}
\label{sec:method-open}

In the survey, we further collected open-ended responses from participants to ask for their detailed rationales for their overall preferences with two questions: (1) Detailed rationales on their valuation process of ranking the explanations with the question, “please briefly describe the rationale on ranking the explanations,” and (2) general perceptions on values of explanations and AI systems, “what is the most important aspect of explanations for you? Do you have any additional comments about the AI system?”

\begin{figure}[h]%
\centering
\includegraphics[width=\columnwidth]{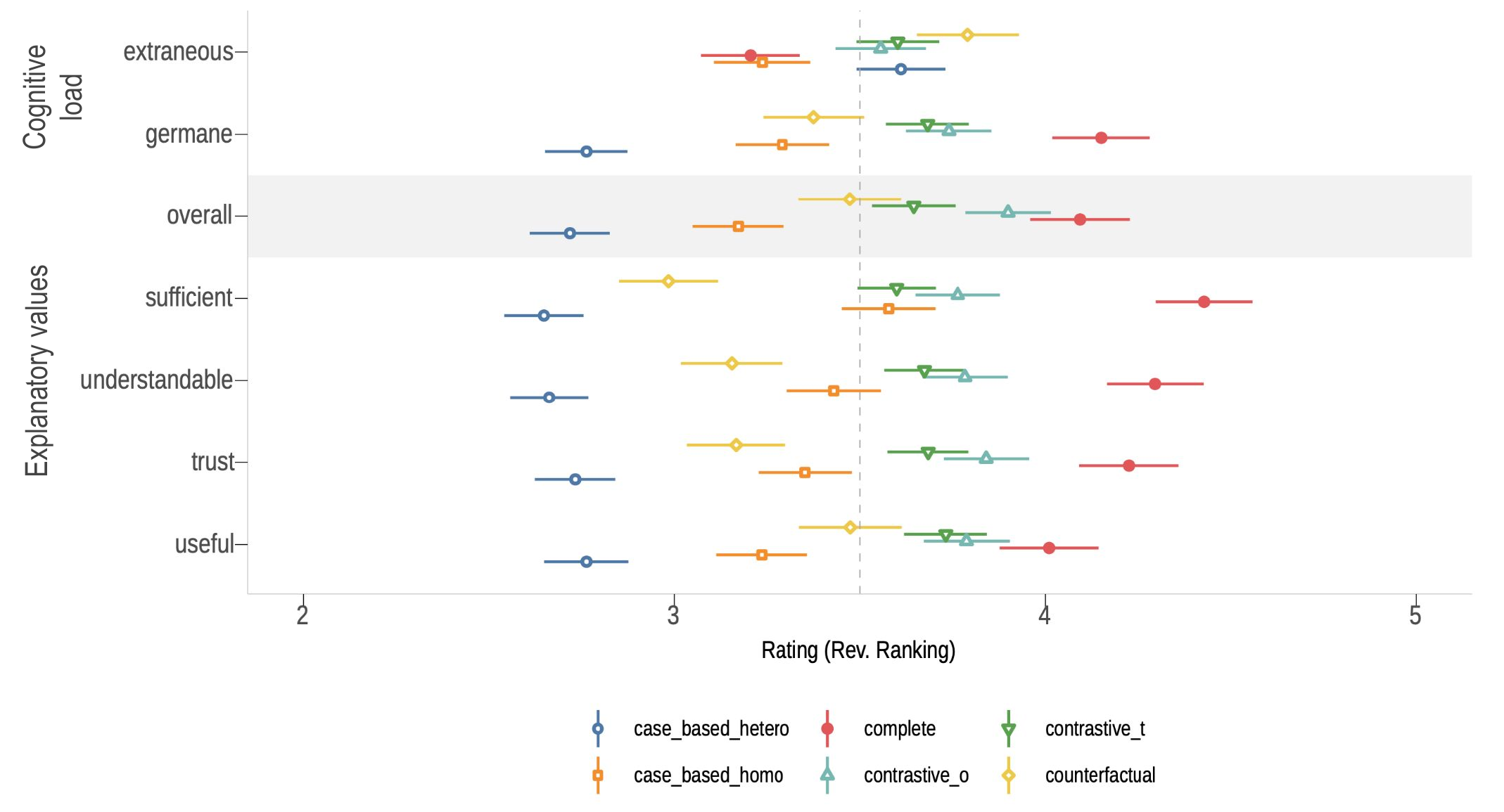}
\caption{The relative ratings (along the $x$-axis) for each of the six explanation variants for the {\it extraneous} and {\it germane} cognitive capacities influenced by how information is presented, as well as the preference rankings in five distinct value dimensions. To facilitate the summary, participants' {\it overall} preference was highlighted in gray.}\label{fig:overallPref}
\end{figure}

To extract and summarize the characteristics of their evaluation process from texts, we took a two-step approach. First, we used qualitative coding to identify recurring themes in participants' valuations. Two coders read a random sample of 30\% of written responses and engaged in multiple rounds of discussion to group them into high-level categories, resulting in four distinct types of valuations: explanatory properties, features, comparison, and XAI system. With these categories, we iterated over all written responses to quantitatively and qualitatively identify whether each written response contains any of four valuation categories (e.g., explanatory values) and what specific values (e.g., simple, easy) are encoded. The definitions and methods of analysis for four categories are summarized below:

\begin{itemize}
    \item \textbf{Property-oriented valuation}: 524 responses mentioned the desired properties of explanations regarding information quantity, logic, or value (e.g., brief, simple, informative). We extracted adjectives that express positive explanatory values by conducting the part-of-text analysis and filtering out negative and irrelevant words.
    
    \item \textbf{Feature-oriented valuation}: 135 responses referred to certain feature(s) given in the decision-making scenario (e.g., credit score, BMI, weather, genre, or age) as a rationale of why they chose certain explanations as the least/most preferred. From these responses, we extract some findings on what aspects of information selectivity mainly influence the valuation of explanations.
    
    \item \textbf{Comparison-related valuation}: 77 responses preferred certain type(s) of contrastive strategies. As this type of feedback was expressed in an unstructured manner, we manually coded the types of comparisons mentioned in their responses.
    \item \textbf{XAI-system-related valuation}
    25 responses expressed their expectations on how explanations in AI systems were received or what they are supposed to do. Similarly, we manually coded whether each response contained system-related valuation.
\end{itemize}

%% file: 050_Results.tex
\section{Results}

\input{051_RQ1}
\input{052_RQ2}
\input{053_RQ3}
\input{054_RQ4}

%% file: 051_RQ1.tex
\subsection{How do participants' perceived value of explanations differ, overall and within sociotechnical contexts?}\label{sec:values}

\begin{figure}[H]%
\centering
\includegraphics[width=\columnwidth]{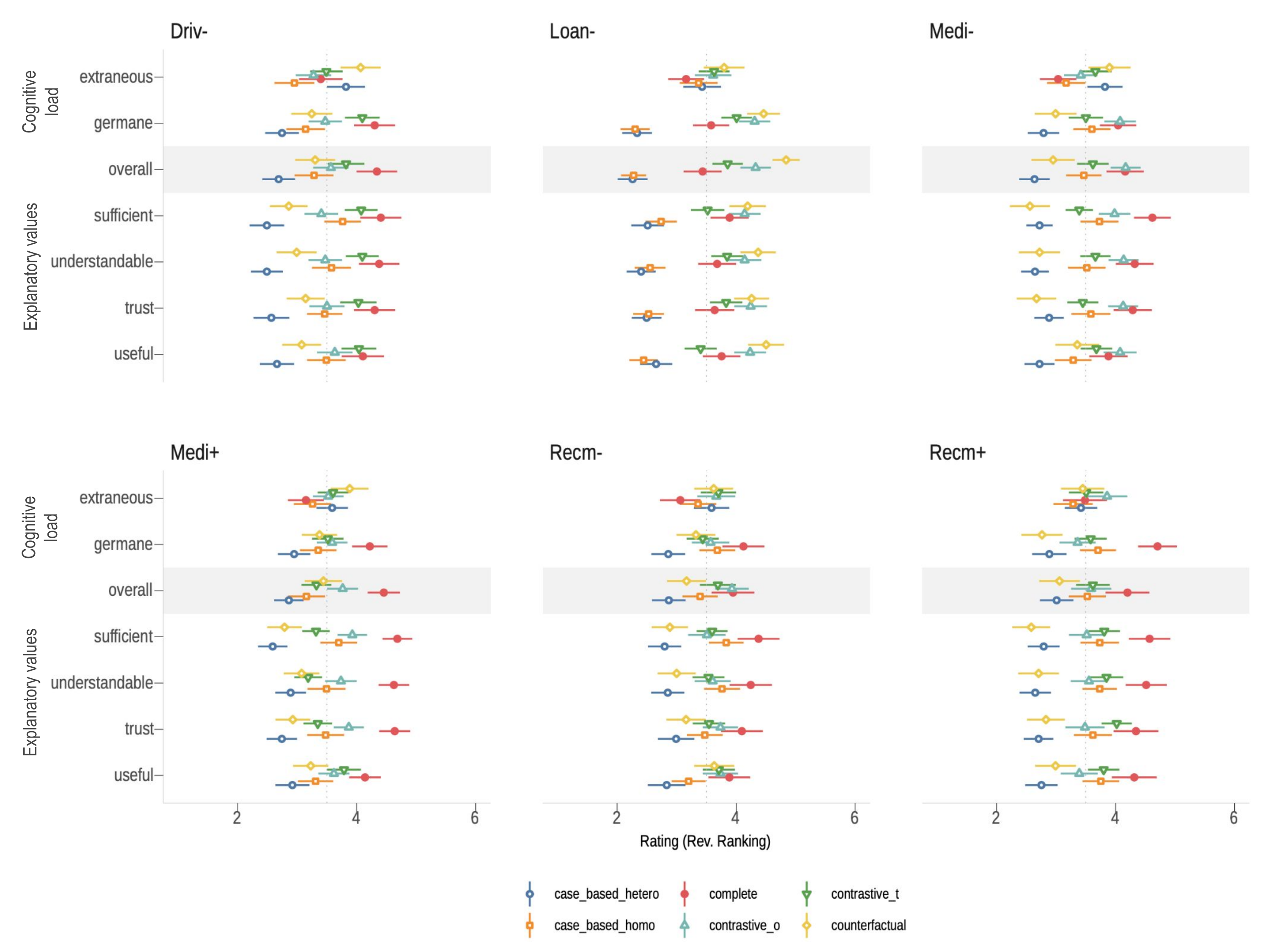}
\vspace{-2em}
\caption{In each scenario, the relative ratings (along the $x$-axis) for each of the six explanation variants for the {\it extraneous} and {\it germane} cognitive capacities, as well as the preference rankings in five distinct value dimensions. To facilitate the summary, participants' {\it overall} preference was highlighted in gray.}\label{fig:overallPref-context}
\end{figure}

First, we examine how explanation strategies were perceived differently in terms of perceived values of explanations overall and within each decision context, which are summarized in Fig.~\ref{fig:overallPref} and Fig.~\ref{fig:overallPref-context} respectively.

Our analysis found \comp explanations were evaluated as higher in all explanatory values than any other explanation types. Conversely, \cbhe was considered the least favorable in all valuation aspects. \cf and \cbho were ranked second-to-worst in terms of {\it sufficient} and {\it understandable}, and {\it trustworthiness} and {\it usefulness} respectively.

However, when analyzing the ratings of explanatory values for each decision scenario, we found that participants’ perceptions of these explanations varied depending on the scenarios they encountered. In high-stakes and time-sensitive scenarios such as \drivN, \loanN, and \mediN, three explanation styles stood out. The \ctt received a rating comparable to the \comp in \drivN. The \ctt explanation, ``\textit{The estimated arrival time was predicted 20 minutes later than usual because there was heavy snow last night ..., while your usual travel with on-time arrival happened without snow with good traffic flow},'' was considered to be a {\it sufficient}, {\it understandable}, {\it trustworthy}, and {\it useful} explanation for the AI-decision ``{\it your arrival time as being 20 minutes later than usual ...}'' This suggests that an explanation contrasting outcomes at different times is particularly applicable to highly time-sensitive situations. In \mediN, the \cto (``...\textit{You are at a high risk for Acute Coronary Syndrome because you have high cholesterol and blood pressure between 90-145. In contrast, a user who wasn't....}'') received ratings comparable to the \comp. This indicates that justifications for the AI's decision based on contrasting results due to different causes work well in medical contexts.

The \loanN scenario exhibited a distinct pattern of valuation compared to all other scenarios. Participants in this scenario favored the \comp less than those in other scenarios. On the other hand, the \cf (``{\it You could have been granted a loan if you had been employed with an annual income more than...}'') as well as \cto explanations were the most popular choices. This suggests that both explanation approaches were particularly useful in a situation where there was either a win or a loss, such as with \loanN.

%% file: 052_RQ2.tex
\subsection{Which aspects and properties of explanations are linked to the distinct valuation of those explanation strategies?}
Based on the analysis of open-ended responses, we examine participants' detailed rationale on what aspects and properties of explanations made them prefer certain explanations over the others. We find evidence on all facets of explanation strategies listed in Table \ref{tab:comp_strategies} including contrastive strategies (how and what/who to compare) and information selectivity (information complexity and alignment) as well as general expectations over AI systems' explainability as follows. 

\subsubsection{Participants appreciate different types of explanatory properties, mostly pertaining to information complexity.}
\label{sec:exp-properties}
First, most of the participants (524/698, 75.1\%) tended to evaluate explanation strategies based on a variety of explanatory properties to further elaborate on their preferences in the quantitative results. The five most frequent properties across all explanation styles were easy (126 times), clear (66), relevant (38), specific (36), and detailed (34). 

\begin{figure}[H]%
\centering
\includegraphics[width=.8\columnwidth]{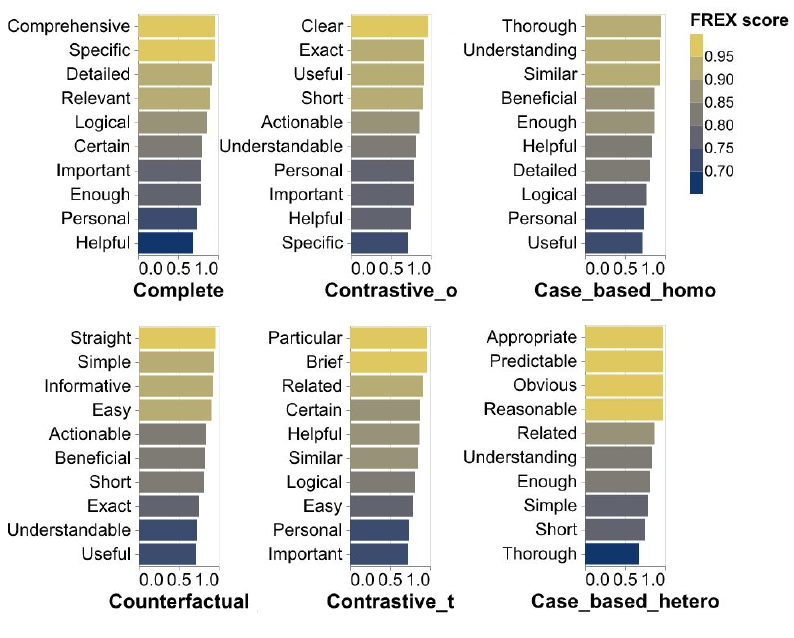}
\caption{The most frequent explanatory properties in association with the explanation strategies.}
\label{fig:exp-properties}
\end{figure}

It was noticeable that the top explanatory properties in each explanation type were mostly pertaining to information complexity such as detailed, simple, or brief. The lists of top-10 most associated properties in favor of each explanation strategy (Fig. \ref{fig:exp-properties}) were extracted based on FREX score (FRequency and EXclusivity) \cite{bischof2012summarizingfrex}, which identifies words that are both frequent in and exclusive to a topic of interest, regarding each explanation style. Specifically, \comp was perceived as comprehensive and detailed as it described ``\textit{every aspect of why the decision was made}'' and the comprehensive information allowed them to``\textit{form a big picture in [their] mind.}.'' \cbho was appreciated for not only being thorough but also its comparative characteristic, ``\textit{being detailed and comparative would help to highlight the urgency of the situation.}'' On the other hand, contrastive explanations (including \cto and \ctt, and \cf) were commonly appreciated by virtue of their clear, brief, and short information representation as it gives them ``\textit{the most exact and particular reason what the key problem is to improve}''.

On the other hand, contrastive/counterfactual explanations were preferred due to both comparative strategies. For example, \cto and \cf were received more actionable because it helps ``\textit{take away something from and apply it to my life.}'' due to being descriptive of their status in contrast to  the ones with the opposite outcome. On the other hand, \ctt explanations contrasting with previous state of the user were regarded as more relevant as it ``\textit{relates to me more and compares less with others.}''

\subsubsection{Participants favor explanations they can readily accept based on their prior knowledge or beliefs.}
\label{sec:selective}
In the 135 responses (19.3\%), participants expressed preferences for explanation styles when the explanation mentioned one or more feature(s) they perceived as critical or unnecessary in the decision-making. 

These feature-oriented valuation, spanning 34 types of features for 205 times, varied across decision contexts but appeared mainly in less knowledge-intensive decision-making contexts. For instance, in the loan approval context, participants often confidently referred to employment (\loanN: 17) or annual income (\loanN: 17) as critical factors of loan approval based on their beliefs or prior experiences as commented, ``\textit{income matters.}'' or ``\textit{not being employed is the most critical problem. There is no other reason that matters.}''. In movie recommendations, the browsing history was often considered as a critical factor (\recomN: 30, \recomP: 23) because this feature ``\textit{made the most sense}'' and they feel ``\textit{this will give me the best choice for movies.}'' The movie genre (\recomN: 9, \recomP: 7) was also often perceived as an understandable and important feature to the given recommendations, ``\textit{the relation to the genre and my previous recommendations make it the best and easiest to understand.}'' On the other hand, the least frequent contexts were medical diagnosis (\mediN: 10, \mediP: 12) with mentions of some features such as blood pressure (\mediN: 2, \mediP: 3) or cholesterol (\mediN: 2, \mediP: 1). This suggests that, while some participants prefer explanations that seem more plausible to them, possibly due to their prior beliefs and knowledge, this preference may vary depending on the level of professional knowledge required for the decision context.

On the other hand, some features, especially those related to demographics, also provoked negative preferences over explanations. A number of participants (\recomN: 15, \recomP: 12, \loanN: 5) considered all demographic-related features (such as gender, marriage, or age) as not useful, necessary, or relevant to the AI's decision. Some of them raised their concerns about privacy or profiling issues of collecting and processing personal data. 

\input{tables/tab_comparative_strategies}

\subsubsection{Rather than simply contrastive, human-friendly explanations depend on whether, what, and how to compare.}
\label{sec:contrastive}
77 participants (11.0\%) showed diverse user preferences over different types of comparisons, based on factors like what or whom is being compared against their own status. In Table \ref{tab:comp_strategies}, explanations are presented based on their average ranking over a group of participants and highlighting liked (red) or disliked (blue) strategies. 

We found that some participants did not want any types of comparisons and preferred \comp over all others (+: \comp; -: all others). They preferred explanations to ``\textit{focus on themselves}'' rather than others because ``\textit{everyone has different personal context}'' but AI may not consider all contexts when making comparisons. Another group of participants did not like being compared to others but rather to their own previous or hypothetical state (+: \cf, \ctt; -: all others). On the other hand, others preferred certain comparative explanations. For example, a group of participants favored comparison with typical or similar others (+: \cbho, \cbhe; -: all others) or those with opposite outcome (+: \cto, \ctt; -: all others).

\subsubsection{Participants have different expectations over AI explainability.}
\label{sec:ai-system}

Participants' expectations about the role of explanations in AI systems varied significantly across the 25 responses (3.5\%). First, some expressed their views on the objectives of XAI systems. Ten participants, for instance, anticipated that AI systems would provide professional information or be used for complex tasks. Notably, eight participants preferred explanations that included more statistics and numbers, citing that such information ``{\it helps [in] visualizing the status.}'' Two participants mentioned that AI systems should present professional information that could provide more learning opportunities, as opposed to merely presenting well-known terms and factors such as BMI or blood pressure.

Second, four participants argued that AI systems should do less, suggesting that the use of XAI should be limited or restricted to human experts. They voiced concerns such as, ``{\it I [would] rather hear this from a human [than AI]; it is useful when used [alongside] doctors}'' and ``{\it AI is just for diagnosing; to tell me whether I should go to a doctor.}''

%% file: tables/tab_comparative_strategies.tex
\addtolength{\tabcolsep}{-4pt}
\begin{table*}[h]
\caption{Types of comparative strategies disliked (red) or liked (blue) by users.}
\label{tab:comp_strategies}
\begin{tabular}{@{}lllllllll@{}}
\toprule
 &
   &
  \multicolumn{6}{l}{\textbf{\begin{tabular}[c]{@{}l@{}}Explanation strategies ranked by\\ preference scores\end{tabular}}} &
   \\ \cmidrule(lr){3-8}
\multirow{-3}{*}{\textbf{\begin{tabular}[c]{@{}l@{}}Types of comparative strategies\\ disliked/liked by users\\ in written responses\end{tabular}}} &
  \multirow{-2}{*}{\textbf{\begin{tabular}[c]{@{}l@{}}\#\\ users\end{tabular}}} &
  \begin{tabular}[c]{@{}l@{}}\textbf{1st}\end{tabular} &
  \textbf{2nd} &
  \textbf{3rd} &
  \textbf{4th} &
  \textbf{5th} &
  \textbf{6th} \\ \midrule
\cellcolor[HTML]{FFCCC9}\textit{"I \uline{don't like} any comparisons."} &
  13 &
  {\color[HTML]{000000} CP} &
  \cellcolor[HTML]{FFCCC9}{\color[HTML]{000000} CT-T} &
  \cellcolor[HTML]{FFCCC9}{\color[HTML]{000000} CT-O} &
  \cellcolor[HTML]{FFCCC9}{\color[HTML]{000000} CF} &
  \cellcolor[HTML]{FFCCC9}{\color[HTML]{000000} CB-HO} &
  \cellcolor[HTML]{FFCCC9}{\color[HTML]{000000} CB-HT} \\ \midrule
\cellcolor[HTML]{FFCCC9}\textit{\begin{tabular}[c]{@{}l@{}}"I \uline{don't like} being compared to \\ others."\end{tabular}} &
  14 &
  {\color[HTML]{000000} CF} &
  {\color[HTML]{000000} CT-T} &
  \cellcolor[HTML]{FFCCC9}{\color[HTML]{000000} CT-O} &
  \cellcolor[HTML]{FFCCC9}{\color[HTML]{000000} CP} &
  \cellcolor[HTML]{FFCCC9}{\color[HTML]{000000} CB-HO} &
  \cellcolor[HTML]{FFCCC9}{\color[HTML]{000000} CB-HT} \\ \midrule
\cellcolor[HTML]{FFCCC9}\textit{\begin{tabular}[c]{@{}l@{}}"I \uline{don't like} being compared to\\ others who are similar to me."\end{tabular}} &
  14 &
  {\color[HTML]{000000} CP} &
  {\color[HTML]{000000} CT-O} &
  {\color[HTML]{000000} CF} &
  \cellcolor[HTML]{FFCCC9}{\color[HTML]{000000} CT-T} &
  \cellcolor[HTML]{FFCCC9}{\color[HTML]{000000} CB-HO} &
  \cellcolor[HTML]{FFCCC9}{\color[HTML]{000000} CB-HT} \\ \midrule
\cellcolor[HTML]{CBCEFB}\textit{\begin{tabular}[c]{@{}l@{}}"I \uline{like} being compared to others \\ who are typical or similar to me."\end{tabular}} &
  17 &
  \cellcolor[HTML]{CBCEFB}{\color[HTML]{000000} CB-HO} &
  \cellcolor[HTML]{CBCEFB}{\color[HTML]{000000} CB-HT} &
  {\color[HTML]{000000} CP} &
  {\color[HTML]{000000} CF} &
  {\color[HTML]{000000} CT-T} &
  {\color[HTML]{000000} CT-O} \\ \midrule
\cellcolor[HTML]{CBCEFB}\textit{\begin{tabular}[c]{@{}l@{}}"I \uline{like} being compared to others \\ who have the opposite outcome."\end{tabular}} &
  6 &
  \cellcolor[HTML]{CBCEFB}{\color[HTML]{000000} CT-O} &
  {\color[HTML]{000000} CF} &
  {\color[HTML]{000000} CP} &
  {\color[HTML]{000000} CT-T} &
  {\color[HTML]{000000} CB-HO} &
  {\color[HTML]{000000} CB-HT} \\ \midrule
\cellcolor[HTML]{CBCEFB}\textit{\begin{tabular}[c]{@{}l@{}}"I \uline{like} being compared to \\ my previous condition."\end{tabular}} &
  4 &
  \cellcolor[HTML]{CBCEFB}{\color[HTML]{000000} CT-T} &
  \cellcolor[HTML]{FFFFFF}{\color[HTML]{000000} CP} &
  {\color[HTML]{000000} CT-O} &
  {\color[HTML]{000000} CB-HO} &
  {\color[HTML]{000000} CF} &
  {\color[HTML]{000000} CB-HT} \\ \bottomrule
\end{tabular}
\end{table*}
\addtolength{\tabcolsep}{4pt}

%% file: 053_RQ3.tex
\subsection{How do individual, contextual, and cognitive factors interact in the valuation process of explanations?} 

In the second part of our study, we conduct an in-depth analysis of how the distinct values of various explanation strategies, as discussed above, are shaped by a range of factors and their complex interactions. We begin by examining the relationships between sociotechnical and cognitive factors and how they interact with the evaluation of explanations.

\textbf{Sociotechnical contexts are perceived as having distinct contextual properties.}\label{sec:scenario-properties} First, we confirmed that the six scenarios in our study exhibited different characteristics along the three dimensions, including {\it high-stakes}, {\it professional}, and {\it timely} based on participants' ratings (Fig.\ref{fig:situ}). For instance, \loanN, \mediN, and \drivN were rated as more {\it high-stakes} and {\it timely} than \recomP and \recomN (Mann-Whitney U test with $p<0.001$ on each pair of the former and latter groups). Similarly, \mediN and \mediP were rated as more {\it professional} contexts than \recomP and \recomN (Mann-Whitney U test with $p<0.001$ on each pair of the former and latter groups). Fig.\ref{fig:situ} in the \appsec{sec:contextual-prop-and-cog-engagement} illustrates the characteristics of the six different decision scenarios and their impact on participants' comprehension processes.

\textbf{Participants' cognitive engagement tend to differ across sociotechnical contexts.}\label{sec:scenario-properties} We also find that, depending on decision contexts with different contextual properties, participants tend to exhibit different level of having motivations, opportunities, and abilities. For instance, participants in more high-stakes scenarios, such as \loanN and \mediN, showed higher {\it motivation} to investigate the information due to perceived benefits or threats from the consequences of the decision, compared to those in less high-stakes scenarios, such as \recomP and \recomN (Mann-Whitney U test with $p<0.001$ on every pair of the former and latter groups).
In the scenarios with more professional knowledge required (\mediP and \mediN), participants generally felt less capable of making sense of the information on their own. This was supported by the Mann-Whitney U test results, showing a significant difference ($p < 0.001$) between \mediP (\mediN) and almost all other scenarios. In a time-sensitive scenario such as \drivN, participants were more likely to perceive a lower {\it opportunity} to evaluate options thoroughly. The average rating for {\it intrinsic} cognitive capacity was greater than 3.5 across all scenarios (higher values indicate more manageable complexity), suggesting that the inherent level of difficulty of the given information was considered manageable in all cases.

\textbf{Participants' cognitive load tend to change with different explanation strategies.}\label{sec:exp-overall} Fig.~\ref{fig:overallPref} shows the relative ratings of each of the six explanation variants for the {\it extraneous} and {\it germane} cognitive load, as well as for the preference rankings in terms of different value aspects. To facilitate comparison, all ranking results were inversely coded such that a higher value indicates positive rating (see Section~\ref{sec:stat} for details).

Overall, explanation styles with higher information complexity (\comp and \cbho) received significantly lower \textit{extraneous} ratings (Wilcoxon signed rank test with $p<0.001$), whereas others with lower complexity received higher ratings on extraneous load. The \cf explanation with the lowest amount of information received a relatively high {\it extraneous} rating than all others. This suggests that presenting a small amount of information made participants feel easier to distinguish between important and unimportant information when presented with explanation styles other than the \comp and \cbho. However, this did not necessarily lead to an enhanced understanding of why you were given the decision (i.e., higher \textit{germane} load). Participants tended to give a lower {\it extraneous} rating but the highest {\it germane} rating to the \comp explanation (the Wilcoxon signed rank test showed a significantly lower {\it extraneous} rating for \comp than for four of the others, as well as a significantly higher {\it germane} rating for \comp than for others, with all $p<0.001$), suggesting that while the \comp explanation is not perceived as effective in how the information is presented, it helps enhance their understanding of the given decision.

%% file: 054_RQ4.tex
\subsection{How do these factors collectively influence the preference on valuation of contrastive and selective explanations?}
\begin{figure*}[t]%
\centering
\includegraphics[width=\textwidth]{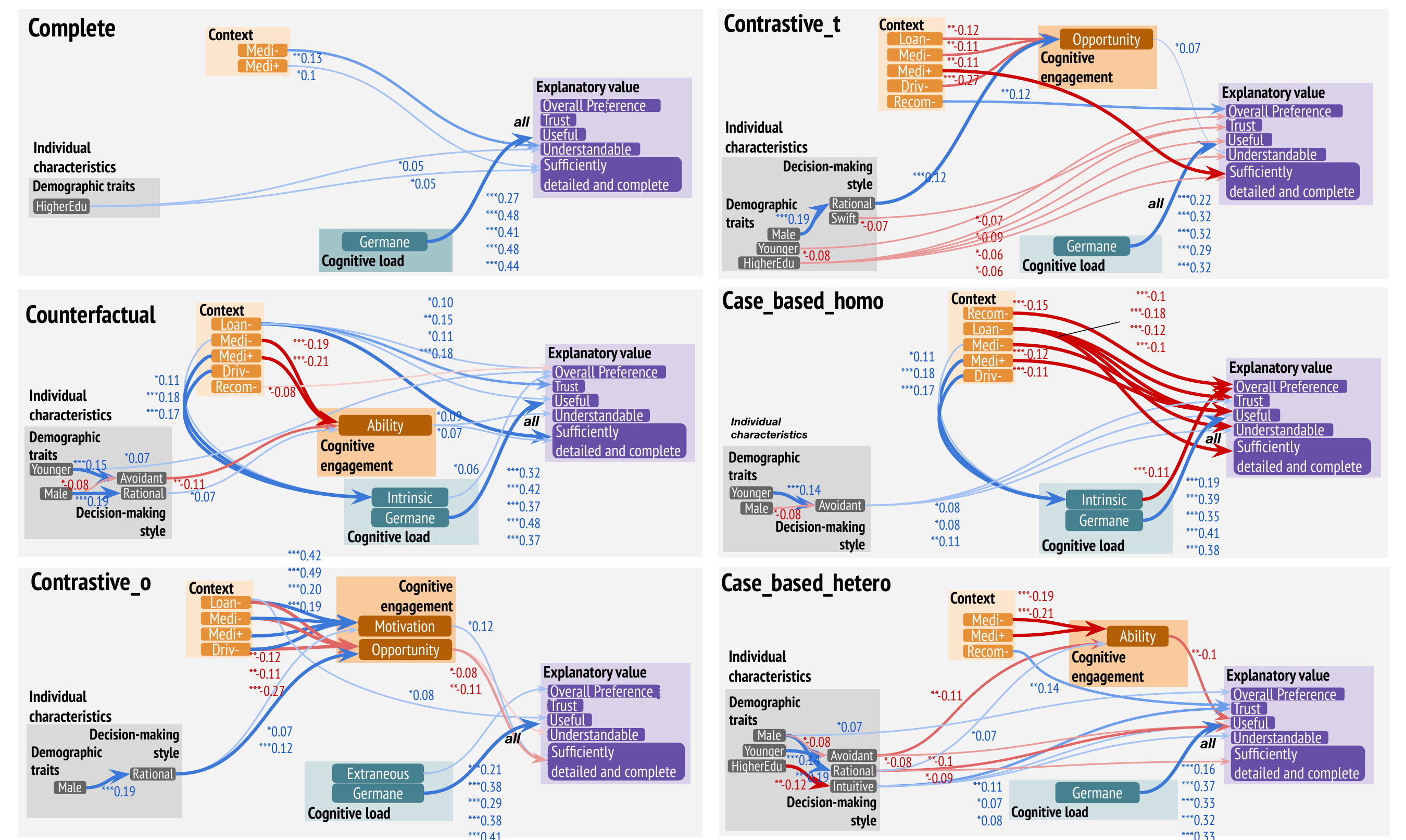}
\caption{Path analysis results for each explanation strategy. Factors significantly influencing the valuation process of each explanation strategy identified in per-strategy path analysis are highlighted.}\label{fig:explanation-analysis}
\end{figure*}

Considering all factors including individual-, context-, and explanation-dependent factors, we investigate the interplay of factors in affecting the valuation of explanations.
As described in Section~\ref{sec:stat}, structural equation models (SEM) are used to test the effects of various factors in explaining whether a participant prefers a particular explanation or not for five explanatory values. Fig.~\ref{fig:explanation-analysis} provides a summary of the significant factors identified by the path analysis for each explanatory strategy. For predicting five explanatory values in each of the explanation styles (e.g., \cbhe), the estimated standardized effects of the significant factors are shown with a 95\% confidence interval.

\textbf{Overview.} Our path analysis reveals a combination of direct and indirect effects of individual-, context-, and explanation-dependent factors on the valuation process. Across the results of path analysis over six strategies, the most impactful factor was \textit{germane} load, directly impacting all explanatory values. Conversely, for \cf and \cbho, \textit{intrinsic} load was significantly higher in high-stakes and professional contexts (\loanN, \mediN, \mediP). 

Sociotechnical contexts also significantly impacted the valuation of explanations either directly or indirectly through cognitive engagement. In those cases, contexts were primarily high-stakes or professional, affecting the explanatory values indirectly by invoking people to perceive higher motivation, lower opportunity, or lower ability (in the high-stakes or highly professional contexts), while also directly influencing valuation due to given contexts. 

For individual characteristics, some demographic traits were significantly associated with decision-making styles (Male $\rightarrow$ Rational (***0.19\footnote{Standardized path coefficient with statistical significance (*p $\leq$ 0.05, **p $\leq$ 0.01, ***p $\leq$ 0.001)}); Younger $\rightarrow$ Dependent(**0.12) and Avoidant (***0.14); HigherEdu $\rightarrow$ Dependent (**0.10) and Intuitive (**-0.12)). The decision-making styles, in turn, tended to influence cognitive engagement. For instance, rational thinkers tended to demonstrate higher motivation (*0.07), opportunity (***0.12), and ability (*0.07), different than avoidant thinkers who tended to have lower ability (**-0.11). These associations together formulate indirect effects of demographic traits towards cognitive engagement via demographic styles. For example, younger individuals, often avoidant, tended to have lower perceived ability. On the other hand, extraneous and germane loads were not significantly associated with individual characteristics.  

\textbf{Strategy-specific characteristics of valuation process.} The SEMs fitted to each explanation strategy provides insights into how the valuation process differs across strategies.

First, \comp explanations' values were mostly shaped by positive direct effects of individual and context-dependent factors. For example, \comp was found more understandable for users with higher education or within medical decision contexts (\mediP, \mediN) requiring advanced professional knowledge. Conversely, \ctt was the counterpart of \comp, as indicated by negative direct associations in a professional context (\mediP $\rightarrow$ Sufficiently-detailed; **-0.11) and positive associations in a low-stakes context (\recomN $\rightarrow$ Preference; **0.12). 

For \cf explanations, both indirect and direct effects of context-dependent factors were notable. \cf was highly valued in \loanN across four explanatory values and in high-stakes contexts due to their high intrinsic load (\mediN, \mediP, \drivN $\rightarrow$ Intrinsic; *0.11, ***0.18, ***0.17, Intrinsic $\rightarrow$ Preference; *0.06), or when individuals have higher perceived ability (Ability $\rightarrow$  Understandable, Sufficiently-detailed; *-0.09, *-0.07). The valuation process of \cbho and \cbhe were the opposite of \cf's. \cbho was less valued directly in \loanN context or indirectly in contexts associated with higher \textit{intrinsic} load (Intrinsic $\rightarrow$ Preference; ***-0.11). \cbhe was positively valued as useful when individuals have lower perceived ability (Ability $\rightarrow$ Useful; ***-0.1).

%% file: 060_Discussion.tex
\section{Discussion}\label{sec:discussion}
Recent research in explainable AI (XAI) has focused on identifying the key features of user-friendly explanations and exploring cognitive frameworks that enhance users' understanding of AI-assisted decisions. In this section, we discuss how our findings both support and challenge existing theories of explanation design. We also provide practical recommendations for designing XAI interfaces, such as AI chatbots, highlighting the need to prioritize interactive and personalized explanation processes tailored to diverse user profiles and decision-making contexts.

\subsection{Effectiveness of Contrastive and Selective explanations}

As discussed in Section \ref{sec:intro}, the properties of human-friendly explanations---especially contrastive and selective explanations---have garnered increasing attention in recent XAI research and are often used as theoretical foundations \cite{SelectiveMutableDialogicXAIReview, SelectiveExplanationsLeveragingHumanInput}. However, our study presents empirical evidence that challenges two key properties in AI-assisted decision-making contexts:

\begin{enumerate}
\item {\bf Contrastive} explanations: In some contexts, contrastive explanations were not perceived as the most understandable or preferred. While earlier studies highlight their intuitive and human-friendly nature, our findings suggest their effectiveness is highly context-dependent (e.g., contrastive explanations were less effective in time-sensitive situations or when comparing outcomes like winning vs. losing). Participant preferences also varied based on the type of contrast (e.g., comparing with others' outcomes or with their own previous state). For example, when contrasting with another individual's outcome (\cto), participants found explanations actionable, whereas comparing their current state to a previous one (\ctt) was perceived as more relevant to personal decision-making. However, as discussed in Section \ref{sec:contrastive}, some participants found \cto explanations less relevant, as they often focused on their own circumstances.
\item {\bf Selective} explanations: Our analysis of open-ended responses revealed that selective explanations should not be assumed universally effective. Participants preferred explanations that highlighted specific features (e.g., annual income, browsing history), especially when those features aligned with their expectations. However, this preference was more pronounced in simpler decision contexts, such as loan approvals or movie recommendations, where professional knowledge is not required. In more complex contexts, like medical decision-making, participants were less likely to express preferences related to selective information, as noted in Section \ref{sec:selective}.
\end{enumerate}

Overall, our findings challenge the assumption that contrastive and selective explanations are universally intuitive or advantageous. These strategies should be carefully tailored to individual needs and contextual factors when incorporated into AI systems.

\subsection{Need to Balance Individual Engagement and Cognitive Load}


Recent XAI literature has introduced cognitive frameworks, such as cognitive forcing \cite{TrustThinkCognitiveForcing} and evaluative AI \cite{miller2023explainable}, aimed at enhancing user engagement and mitigating the risk of over-reliance. These studies suggest using cognitive interfaces that encourage slow, deliberate thinking \cite{kahneman2011thinking} or offer multiple explanations with contrasting outcomes, allowing users to generate their own hypotheses rather than relying solely on system-driven recommendations. While these approaches can improve engagement, they also risk increasing cognitive load, particularly for users inclined toward intuitive thinking. It is unrealistic to expect participants who typically overlook system-generated explanations to actively engage with more complex explainability tasks. Our findings suggest that these approaches should be tailored to individual levels of engagement rather than implemented as one-size-fits-all solutions.

\subsection{The Role of Content and Tone in Shaping User Perceptions of AI Explanations}


Our research revealed that the design of explanations significantly affects participants' perceptions. While most explanations were well received, 15 participants expressed dissatisfaction, particularly with the use of demographic data. One participant noted, ``{\it [Anything] involving demographics or comparisons raises questions of profiling,}'' while another expressed distrust in the system, stating, ``{\it [The] data they collect on me makes me distrust the system.}'' These responses indicate that incorporating personal or comparative data can cause discomfort and raise concerns about privacy and fairness in AI systems.

Additionally, the tone of explanations emerged as an important factor in shaping user perceptions. Six participants stressed that the explanations should not sound ``{\it cold or sarcastic.}'' They preferred AI systems that communicated in a polite, conversational manner, suggesting that the emotional tone is as crucial as the content itself. While our study primarily focused on clarity and utility, these findings highlight the importance of incorporating human-like warmth into AI explanations, particularly in customer support scenarios, as noted in prior research \cite{UnderstandingBenefitsChallengesDeployingConversationala}.

\subsection{Design implications for AI chatbot interfaces}

With the growing adoption of chat-based AI services and large language models, chatbot interfaces have the potential to provide more interactive explanations for AI-driven decisions. Our study highlights several design recommendations for these systems:

\begin{itemize}
    \item {\bf Personalized explanations:} Explanations should be tailored to individual users based on their cognitive traits and contextual factors. This can be achieved by gathering information about the user's engagement, knowledge, and circumstances through interactive prompts. Where sufficient data is available, a data-driven approach can infer user preferences and cognitive states, though it is crucial to explain how these predictions are made and allow users to correct any inaccuracies.
\item {\bf Managing prior knowledge:} To prevent users from favoring explanations that reinforce incorrect prior knowledge (Section \ref{sec:selective}), the system should enable users to explore specific parts of the explanation. Features like question icons or clickable sections can provide further clarification (e.g., ``What does this feature mean?'' or ``Why is this feature important?'').
\item {\bf Interactive explanation customization:} Systems should offer users options to customize the style of explanations they receive. Users could select between simple or detailed explanations, or those that emphasize values such as trustworthiness, clarity, or privacy (Section \ref{sec:exp-properties}). This interactivity allows users to better control their experience and improve the relevance of the information provided.
\end{itemize}

\subsection{Limitations and Future Directions}

This section outlines the study's limitations and potential future research across four key areas: application domains, explanation aversion, understandability measures, and explanation quality.

\paragraph{Application domains}
Our research focused on simulating AI-assisted decision-making for human data subjects, where explanations provide a rationale for the AI's actions. However, the broader field of AI-assisted decision-making encompasses a wide variety of tasks, users, and interactions. Previous studies have examined tasks such as sentiment analysis, content classification \cite{TrustThinkCognitiveForcing,HumanAIInteractionHealthcareThree}, and maze-solving \cite{ExplanationsCanReduceOverrelianceAI}, typically involving laypeople making decisions with AI assistance. In contrast, other studies have explored professional settings like medical diagnosis \cite{DesigningAITrustCollaborationTimeConstrained, ExplainableAIDeadLongLive}, where AI supports clinicians in diagnosing diseases. While our findings on individual differences in valuing explanations may apply to these diverse tasks, other domains may involve different human roles or AI-human collaboration dynamics, potentially altering the effectiveness of various explanation strategies. Future research should consider the unique requirements and challenges of specific application domains to refine explanation strategies accordingly.

\paragraph{Algorithm and explanation aversion}
Our survey did not specifically examine the phenomenon of algorithm or explanation aversion \cite{logg:2019algorithm, dietvorst:2015algorithm}, where users exhibit a general distrust of AI systems or their explanations. Instead, we focused on assessing preferences for different explanation strategies. Only three participants expressed explicit distrust, with comments such as, "{\it I don't trust or respect its explanations}" and "{\it AI should not be diagnosing.}" However, due to the design of our study, we were unable to explore these negative perceptions in detail. Future research should delve deeper into these attitudes, possibly by comparing user preferences for receiving explanations versus none. Gaining a clearer understanding of the cognitive and contextual factors that contribute to explanation or algorithm aversion could help in designing more effective and trustworthy AI systems.

\paragraph{Measuring perceived and actual understandability}
While our study focused on evaluating the perceived effectiveness of various explanation strategies, subjective perceptions alone may not fully capture the depth of participants' understanding. Perceived understandability is important, as it influences user satisfaction and trust, but it doesn't necessarily indicate how well users comprehend AI-driven decisions. Future research may incorporate objective metrics---such as task accuracy, decision-making speed, and users' ability to make informed decisions based on explanations---to better gauge actual understanding. By combining subjective and objective measures, researchers can gain a holistic view of how explanation strategies impact both user confidence and cognitive comprehension. This would provide more robust insights into the design of AI systems that not only explain decisions clearly but also empower users to act on them effectively. Improving both perceived and actual understanding will be crucial in building trustworthy, reliable AI systems that meet user needs.

\paragraph{Explanation quality}
In our study, we assumed that the explanations provided were accurate and trustworthy. However, there is always a risk that explanations could be designed to mislead or manipulate users. To address this concern, future XAI systems should prioritize transparency, offering users information about uncertainties or the data used to generate the explanations. Investigating user vulnerability to misleading explanations could be an important area of future research. This would involve exploring how users react to inaccurate or biased explanations and developing safeguards to protect against manipulation.

%% file: 070_Conclusion.tex
\section{Conclusion}\label{sec:conclusion}

Previous research on explanation strategies has yielded mixed results, leaving gaps in understanding how individuals value different types of explanations. Drawing from cognitive and social science literature, our study aimed to explore the specific factors influencing user preferences for various explanation strategies. Using a scenario-based survey and a between-subject experiment, we investigated the relationships between explanation types, human valuations, and contextual factors.

Our findings challenge the assumption that human-friendly explanations, such as selective and contrastive strategies, are always beneficial. We showed that complete explanations were generally preferred across most values, while selective explanations were favored primarily in less professional contexts. Additionally, participants showed diverse preferences for comparative strategies, instead of the contrastive approach typically emphasized in prior studies.

We identified several key factors---demographics, cognitive load, and contextual conditions---that significantly shape individual preferences for explanation and comparative strategies. These results suggest that one-size-fits-all cognitive frameworks, including those promoting deliberate thinking, may fail to effectively engage users with varying cognitive abilities and needs.

As AI systems increasingly influence daily decision-making, there is an urgent need to personalize explainability according to users' individual contexts and cognitive traits. Our study offers actionable insights for designing AI systems that deliver more interactive and tailored explanations. By accommodating user diversity, these systems can enhance user engagement, foster trust, and make AI technologies more accessible to a broader audience.

%% file: 080_appendix.tex


\clearpage
\section{Demographic breakdown}\label{sec:demo-breaks}

\begin{figure}[H]%
\centering
\includegraphics[width=0.5\columnwidth]{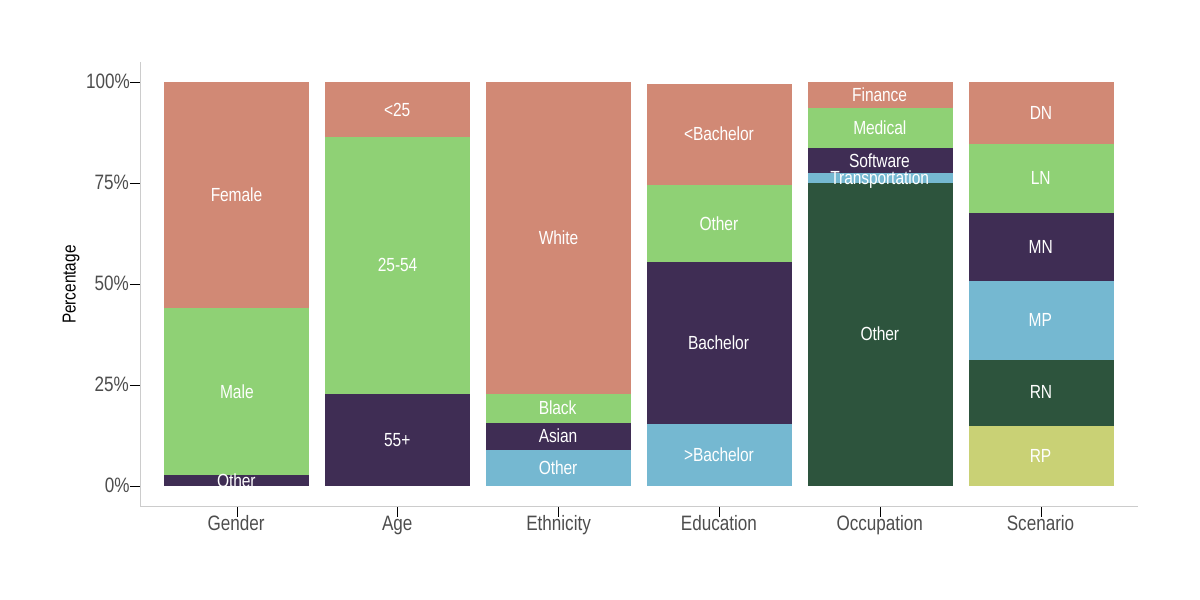}
\vspace{-1em}
\caption{Demographics of participants.}\label{fig:demographics}
\vspace{-1em}
\end{figure}

We collected 839 responses to the survey using the Prolific crowdsourcing platform. Participants had to be 18 or older, fluent in English, and reside in the United States at the time of the survey to ensure the study's validity. To control the quality of these responses, we filtered out those that met one of the exclusion criteria: (1) all five ranking answers (including overall preference and four explanatory values) are identical; (2) the ranking answer for overall preferences was only modified once from its default ranking. This procedure yielded 698 responses as our analysis pool. Based on four demographic questions in the survey, we found that participants were fairly distributed by gender (female: 55.1\%, male: 42.4\%) and age (younger (18-54): 76.7\%, older ($\ge$ 55): 23.4\%).  The majority of the population had a bachelor's degree or higher (55.1\%), and they were the majority across four racial and ethnic groups (White: 77.0\%, Asian: 7.1\%, Black: 6.8\%, Hispanic: 6.6\%, Others: 2.4\%).



\section{Distinct characteristics of six decision scenarios and participants' different level of cognitive engagement}\label{sec:contextual-prop-and-cog-engagement}
\begin{figure}[h]%
\centering
\includegraphics[width=0.5\columnwidth]{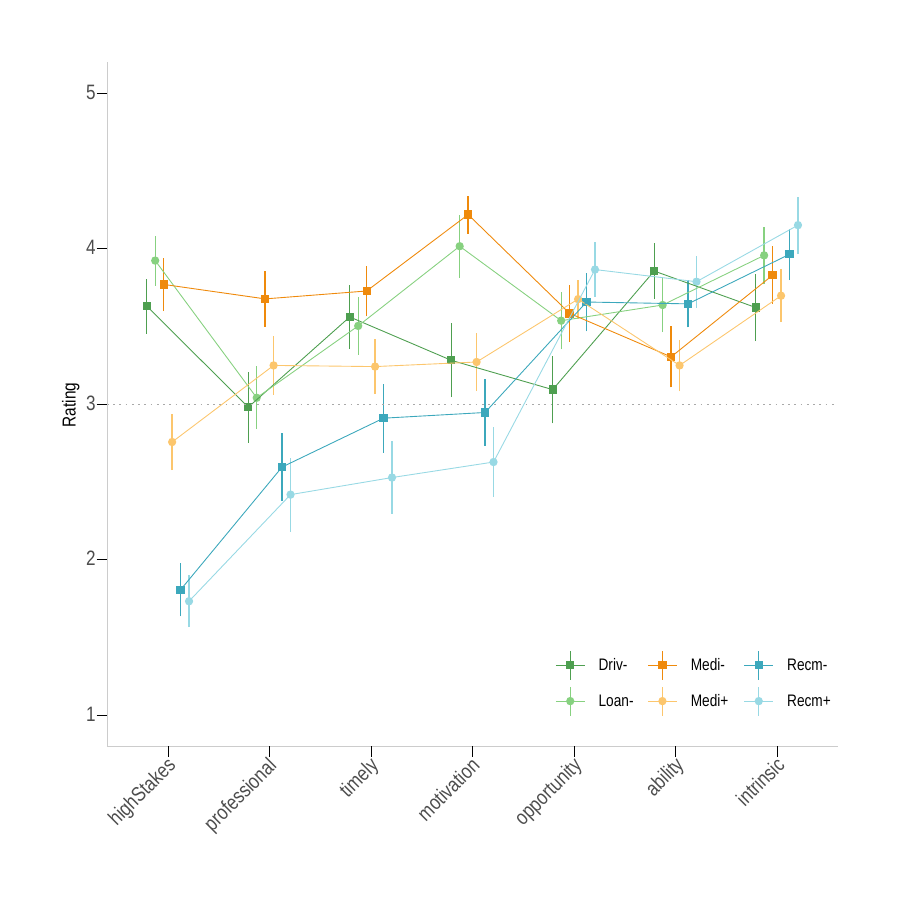}
\vspace{-2em}
\caption{Decision scenarios have distinct characteristics ({\it high-stakes}, {\it professional}, and {\it timely}) that influence participants' comprehending processes, as measured by three MOA variables (individuals' motivation, opportunity, and ability) and the {\it intrinsic} cognitive load (the level of information difficulty covered in the explanations).}\label{fig:situ}
\vspace{-1.5em}
\end{figure}



\section{Survey material}\label{sec:survey}

\begin{figure}[h]%
\centering
\includegraphics[width=\columnwidth]{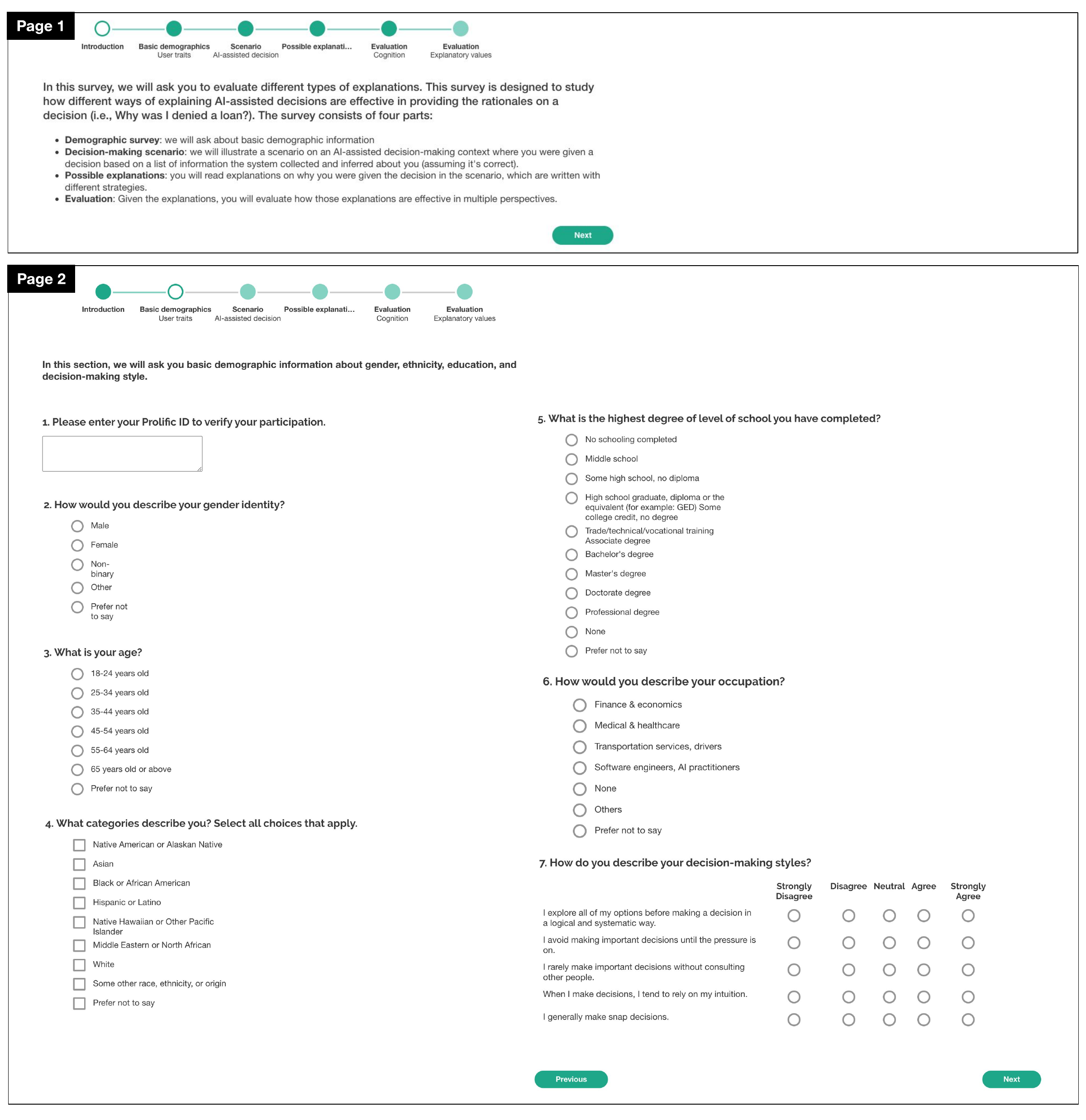}
\caption{Introduction and basic demographics page in the survey.}\label{fig:survey1}
\end{figure}

\begin{figure}[h]%
\includegraphics[width=\columnwidth]{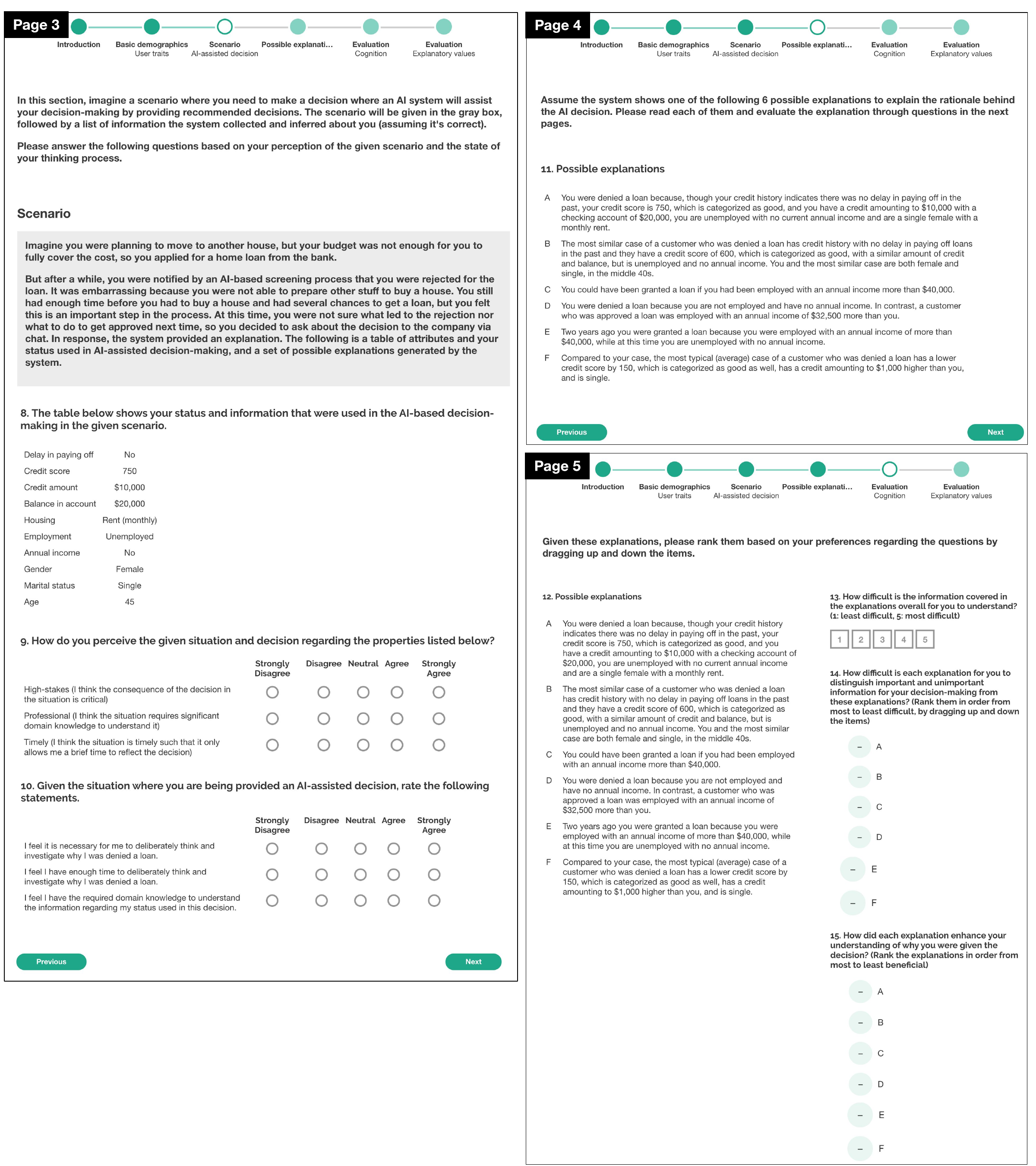}
\caption{Scenario, explanations, and cognition page in the survey.}\label{fig:survey2}
\end{figure}

\begin{figure}[h]%
\centering
\includegraphics[width=\columnwidth]{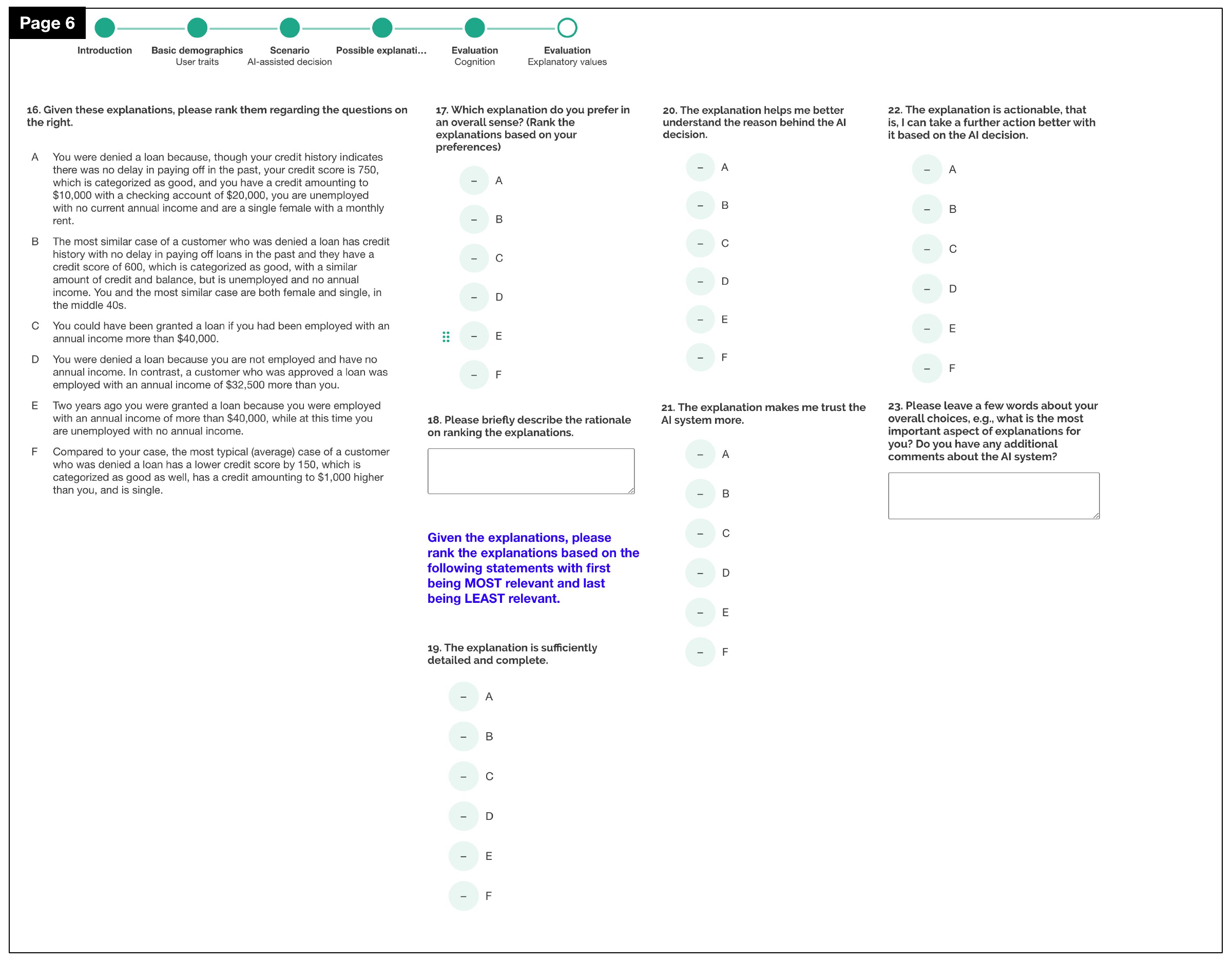}
\caption{Questions on explanatory value in the survey.}\label{fig:survey3}
\end{figure}

\clearpage
\onecolumn
\section{Definitions of Individual/Cognitive Variables}
\addtolength{\tabcolsep}{2pt}
\begin{table*}[h]
\caption{Variables in cognitive load, decision styles, and Motivation-Opportunities-Ability (MOA) model}
\label{tab:cog_vars}
\centering
\begin{NiceTabular}[t]{@{}p{2cm}ll@{}}
\CodeBefore 
\Body
\toprule
\textbf{Variable}                                          & \textbf{Definition}                                                                                                                                                              & \textbf{Scale}                                                                                                                                                                                    \\ \midrule
\multicolumn{3}{l}{\textbf{Cognitive Load Variables} \cite{CognitiveLoadTheoryFormatInstruction,Workingmemorylimitedimprovingknowledge, CognitiveArchitectureInstructionalDesign}}                                                                                                                                                                                                                                                                                                                                                                                             \\ \midrule
\begin{tabular}[t]{@{}l@{}}Intrinsic load\end{tabular}  & \begin{tabular}[t]{@{}l@{}}The inherent level of difficulty associated with \\a specific topic or context\end{tabular}                                                    & \begin{tabular}[t]{@{}l@{}}How difficult is the information covered\\ in the explanations overall for you to understand?\end{tabular}                                                          \\[1.5em]
\begin{tabular}[t]
{@{}l@{}}Extraneous load\end{tabular} & \begin{tabular}[t]{@{}l@{}}Ineffective load by the manner in which information \\ is presented to individuals and is under the control of \\ material\end{tabular} & \begin{tabular}[t]{@{}l@{}}How difficult is each explanation for you \\ to distinguish important and unimportant \\ information for your decision-making from these \\ explanations?\end{tabular} \\[4em]
\begin{tabular}[t]{@{}l@{}}Germane load\end{tabular}    & \begin{tabular}[t]{@{}l@{}}Effective cognitive load devoted to integrating new \\ information, the creation and modification of schema\end{tabular}                           & \begin{tabular}[t]{@{}l@{}}How did each explanation enhance your \\ understanding of why you were given the decision?\end{tabular}                                                             \\ \midrule
\multicolumn{3}{l}{\textbf{Decision Style Variables } \cite{DecisionMakingStyleDevelopmentAssessmentNew}}                                                                                                                                                                                                                                                                                                                                                                                             \\ \midrule
Rational                                                   & \begin{tabular}[t]{@{}l@{}}A thorough search for and logical evaluation of alternatives\end{tabular}                                                                          & \begin{tabular}[t]{@{}l@{}}I make decisions in a logical and systematic way\end{tabular}                                                                                                       \\[0.5em]
Intuitive & A reliance on hunches and feelings & \begin{tabular}[t]{@{}l@{}}When I make decisions, I tend to rely on my intuition\end{tabular}     \\[0.5em]
Avoidant                                                   & An attempt to avoid decision making                                                                                                                                              & \begin{tabular}[t]{@{}l@{}}I avoid making important decisions until the pressure\\ is on\end{tabular}                                                                                            \\[1.5em]
Dependent                                                  & \begin{tabular}[t]{@{}l@{}}A search for advice and direction from others\end{tabular}                                                                                         & \begin{tabular}[t]{@{}l@{}}I rarely make important decisions without consulting \\ other people\end{tabular}                                                                                      \\[1.5em]
Spontaneous                                                & \begin{tabular}[t]{@{}l@{}}A sense of immediacy and a desire to get through the \\ decision-making process as soon as possible\end{tabular}                                   & I generally make snap decisions                                                                                                                                                                   \\ \midrule
\multicolumn{3}{l}{\textbf{Motivation-Opportunities-Ability (MOA) model} \cite{EnhancingMeasuringConsumersMotivationOpportunity}}                                                                                                                   \\ \midrule
Motivation                                                 & \begin{tabular}[t]{@{}l@{}}Conscious and unconscious cognitive processes that \\ directand inspire behavior\end{tabular}                                                      & \begin{tabular}[t]{@{}l@{}}I feel it is necessary for me to deliberately think and \\ investigate the rationale behind the decision.\end{tabular}                                              \\[1.5em]
Opportunity                                                & \begin{tabular}[t]{@{}l@{}}External factors that make a behavior possible\end{tabular}                                                                                        & \begin{tabular}[t]{@{}l@{}}I feel I have enough time to deliberately think and \\ investigate the rationale behind the decision.\end{tabular}                                                  \\[1.5em]
Ability                                                    & \begin{tabular}[t]{@{}l@{}}Individual’s psychological and physical ability to participate\\ in an activity\end{tabular}                                                       & \begin{tabular}[t]{@{}l@{}}I feel I have the required domain knowledge to \\ understand the information regarding my status \\ used in this decision.\end{tabular}                                \\ \bottomrule
\end{NiceTabular}
\begin{tablenotes}
  \small
  \item *All variables were measured using 5-point Likert scale.
\end{tablenotes}
\vspace{-1em}
\end{table*}



%% file: main.bbl

\begin{thebibliography}{81}


\ifx \showCODEN    \undefined \def \showCODEN     #1{\unskip}     \fi
\ifx \showDOI      \undefined \def \showDOI       #1{#1}\fi
\ifx \showISBNx    \undefined \def \showISBNx     #1{\unskip}     \fi
\ifx \showISBNxiii \undefined \def \showISBNxiii  #1{\unskip}     \fi
\ifx \showISSN     \undefined \def \showISSN      #1{\unskip}     \fi
\ifx \showLCCN     \undefined \def \showLCCN      #1{\unskip}     \fi
\ifx \shownote     \undefined \def \shownote      #1{#1}          \fi
\ifx \showarticletitle \undefined \def \showarticletitle #1{#1}   \fi
\ifx \showURL      \undefined \def \showURL       {\relax}        \fi
\providecommand\bibfield[2]{#2}
\providecommand\bibinfo[2]{#2}
\providecommand\natexlab[1]{#1}
\providecommand\showeprint[2][]{arXiv:#2}

\bibitem[Adhikari et~al\mbox{.}(2019)]%
        {adhikari2019leafage}
\bibfield{author}{\bibinfo{person}{Ajaya Adhikari}, \bibinfo{person}{David~MJ Tax}, \bibinfo{person}{Riccardo Satta}, {and} \bibinfo{person}{Matthias Faeth}.} \bibinfo{year}{2019}\natexlab{}.
\newblock \showarticletitle{LEAFAGE: Example-based and Feature importance-based Explanations for Black-box ML models}. In \bibinfo{booktitle}{\emph{2019 IEEE international conference on fuzzy systems (FUZZ-IEEE)}}. IEEE, \bibinfo{pages}{1--7}.
\newblock


\bibitem[Alacreu-Crespo et~al\mbox{.}({[n.\,d.]})]%
        {SpanishvalidationGeneralDecisionMakingStyle}
\bibfield{author}{\bibinfo{person}{Adrián Alacreu-Crespo}, \bibinfo{person}{María~C Fuentes}, \bibinfo{person}{Diana Abad-Tortosa}, \bibinfo{person}{Irene Cano-Lopez}, \bibinfo{person}{Esperanza González}, {and} \bibinfo{person}{Miguel~Ángel Serrano}.} \bibinfo{year}{[n.\,d.]}\natexlab{}.
\newblock \showarticletitle{Spanish validation of {General} {Decision}-{Making} {Style} scale: {Sex} invariance, sex diﬀerences and relationships with personality and coping styles}.
\newblock \bibinfo{journal}{\emph{Judgment and Decision Making}} (\bibinfo{year}{[n.\,d.]}), \bibinfo{pages}{13}.
\newblock


\bibitem[Anjomshoae et~al\mbox{.}(2019)]%
        {ExplainableAgentsRobotsResults}
\bibfield{author}{\bibinfo{person}{Sule Anjomshoae}, \bibinfo{person}{Amro Najjar}, \bibinfo{person}{Davide Calvaresi}, {and} \bibinfo{person}{Kary Främling}.} \bibinfo{year}{2019}\natexlab{}.
\newblock \showarticletitle{Explainable {Agents} and {Robots}: {Results} from a {Systematic} {Literature} {Review}}.
\newblock  (\bibinfo{year}{2019}), \bibinfo{pages}{11}.
\newblock


\bibitem[Bertrand et~al\mbox{.}(2023)]%
        {SelectiveMutableDialogicXAIReview}
\bibfield{author}{\bibinfo{person}{Astrid Bertrand}, \bibinfo{person}{Tiphaine Viard}, \bibinfo{person}{Rafik Belloum}, \bibinfo{person}{James~R. Eagan}, {and} \bibinfo{person}{Winston Maxwell}.} \bibinfo{year}{2023}\natexlab{}.
\newblock \showarticletitle{On {Selective}, {Mutable} and {Dialogic} {XAI}: a {Review} of {What} {Users} {Say} about {Different} {Types} of {Interactive} {Explanations}}. In \bibinfo{booktitle}{\emph{Proceedings of the 2023 {CHI} {Conference} on {Human} {Factors} in {Computing} {Systems}}}. \bibinfo{publisher}{ACM}, \bibinfo{address}{Hamburg Germany}, \bibinfo{pages}{1--21}.
\newblock
\showISBNx{978-1-4503-9421-5}
\urldef\tempurl%
\url{https://doi.org/10.1145/3544548.3581314}
\showDOI{\tempurl}


\bibitem[Bischof and Airoldi(2012)]%
        {bischof2012summarizingfrex}
\bibfield{author}{\bibinfo{person}{Jonathan Bischof} {and} \bibinfo{person}{Edoardo~M Airoldi}.} \bibinfo{year}{2012}\natexlab{}.
\newblock \showarticletitle{Summarizing topical content with word frequency and exclusivity}. In \bibinfo{booktitle}{\emph{Proceedings of the 29th International Conference on Machine Learning (ICML-12)}}. \bibinfo{pages}{201--208}.
\newblock


\bibitem[Bruine~de Bruin et~al\mbox{.}(2007)]%
        {Individualdifferencesadultdecisionmakingcompetence}
\bibfield{author}{\bibinfo{person}{Wändi Bruine~de Bruin}, \bibinfo{person}{Andrew~M. Parker}, {and} \bibinfo{person}{Baruch Fischhoff}.} \bibinfo{year}{2007}\natexlab{}.
\newblock \showarticletitle{Individual differences in adult decision-making competence.}
\newblock \bibinfo{journal}{\emph{Journal of Personality and Social Psychology}} \bibinfo{volume}{92}, \bibinfo{number}{5} (\bibinfo{date}{May} \bibinfo{year}{2007}), \bibinfo{pages}{938--956}.
\newblock
\showISSN{1939-1315, 0022-3514}
\urldef\tempurl%
\url{https://doi.org/10.1037/0022-3514.92.5.938}
\showDOI{\tempurl}


\bibitem[Bussone et~al\mbox{.}(2015)]%
        {RoleExplanationsTrustReliance}
\bibfield{author}{\bibinfo{person}{Adrian Bussone}, \bibinfo{person}{Simone Stumpf}, {and} \bibinfo{person}{Dympna O'Sullivan}.} \bibinfo{year}{2015}\natexlab{}.
\newblock \showarticletitle{The {Role} of {Explanations} on {Trust} and {Reliance} in {Clinical} {Decision} {Support} {Systems}}. In \bibinfo{booktitle}{\emph{2015 {International} {Conference} on {Healthcare} {Informatics}}}. \bibinfo{pages}{160--169}.
\newblock
\urldef\tempurl%
\url{https://doi.org/10.1109/ICHI.2015.26}
\showDOI{\tempurl}


\bibitem[Buçinca et~al\mbox{.}({[n.\,d.]})]%
        {TrustThinkCognitiveForcing}
\bibfield{author}{\bibinfo{person}{Zana Buçinca}, \bibinfo{person}{Maja~Barbara Malaya}, {and} \bibinfo{person}{Krzysztof~Z Gajos}.} \bibinfo{year}{[n.\,d.]}\natexlab{}.
\newblock \showarticletitle{To {Trust} or to {Think}: {Cognitive} {Forcing} {Functions} {Can} {Reduce} {Overreliance} on {AI} in {AI}-assisted {Decision}-making}.
\newblock   \bibinfo{volume}{5} (\bibinfo{year}{[n.\,d.]}), \bibinfo{pages}{21}.
\newblock


\bibitem[Chandler and Sweller(1991)]%
        {CognitiveLoadTheoryFormatInstruction}
\bibfield{author}{\bibinfo{person}{Paul Chandler} {and} \bibinfo{person}{John Sweller}.} \bibinfo{year}{1991}\natexlab{}.
\newblock \showarticletitle{Cognitive {Load} {Theory} and the {Format} of {Instruction}}.
\newblock \bibinfo{journal}{\emph{Cognition and Instruction}} \bibinfo{volume}{8}, \bibinfo{number}{4} (\bibinfo{date}{Dec.} \bibinfo{year}{1991}), \bibinfo{pages}{293--332}.
\newblock
\showISSN{0737-0008, 1532-690X}
\urldef\tempurl%
\url{https://doi.org/10.1207/s1532690xci0804_2}
\showDOI{\tempurl}


\bibitem[Chiyah~Garcia et~al\mbox{.}(2018)]%
        {ExplainableAutonomyStudyExplanation}
\bibfield{author}{\bibinfo{person}{Francisco~Javier Chiyah~Garcia}, \bibinfo{person}{David~A. Robb}, \bibinfo{person}{Xingkun Liu}, \bibinfo{person}{Atanas Laskov}, \bibinfo{person}{Pedro Patron}, {and} \bibinfo{person}{Helen Hastie}.} \bibinfo{year}{2018}\natexlab{}.
\newblock \showarticletitle{Explainable {Autonomy}: {A} {Study} of {Explanation} {Styles} for {Building} {Clear} {Mental} {Models}}. In \bibinfo{booktitle}{\emph{Proceedings of the 11th {International} {Conference} on {Natural} {Language} {Generation}}}. \bibinfo{publisher}{Association for Computational Linguistics}, \bibinfo{address}{Tilburg University, The Netherlands}, \bibinfo{pages}{99--108}.
\newblock
\urldef\tempurl%
\url{https://doi.org/10.18653/v1/W18-6511}
\showDOI{\tempurl}


\bibitem[Chromik and Schuessler(2020)]%
        {TaxonomyHumanSubjectEvaluation}
\bibfield{author}{\bibinfo{person}{Michael Chromik} {and} \bibinfo{person}{Martin Schuessler}.} \bibinfo{year}{2020}\natexlab{}.
\newblock \showarticletitle{A {Taxonomy} for {Human} {Subject} {Evaluation} of {Black}-{Box} {Explanations} in {XAI}}.
\newblock  (\bibinfo{year}{2020}), \bibinfo{pages}{7}.
\newblock


\bibitem[Cunningham et~al\mbox{.}(2003a)]%
        {EvaluationUsefulnessCaseBasedExplanation}
\bibfield{author}{\bibinfo{person}{Pádraig Cunningham}, \bibinfo{person}{Dónal Doyle}, {and} \bibinfo{person}{John Loughrey}.} \bibinfo{year}{2003}\natexlab{a}.
\newblock \showarticletitle{An {Evaluation} of the {Usefulness} of {Case}-{Based} {Explanation}}.
\newblock In \bibinfo{booktitle}{\emph{Case-{Based} {Reasoning} {Research} and {Development}}}, \bibfield{editor}{\bibinfo{person}{Kevin~D. Ashley} {and} \bibinfo{person}{Derek~G. Bridge}} (Eds.). Vol.~\bibinfo{volume}{2689}. \bibinfo{publisher}{Springer Berlin Heidelberg}, \bibinfo{address}{Berlin, Heidelberg}, \bibinfo{pages}{122--130}.
\newblock
\showISBNx{978-3-540-40433-0}
\urldef\tempurl%
\url{https://doi.org/10.1007/3-540-45006-8_12}
\showDOI{\tempurl}
\newblock
\shownote{Series Title: Lecture Notes in Computer Science}.


\bibitem[Cunningham et~al\mbox{.}(2003b)]%
        {EvaluationUsefulnessCaseBasedExplanationa}
\bibfield{author}{\bibinfo{person}{Padraig Cunningham}, \bibinfo{person}{Dónal Doyle}, {and} \bibinfo{person}{John Loughrey}.} \bibinfo{year}{2003}\natexlab{b}.
\newblock \bibinfo{booktitle}{\emph{An {Evaluation} of the {Usefulness} of {Case}-{Based} {Explanation}}}.
\newblock
\urldef\tempurl%
\url{https://doi.org/10.1007/3-540-45006-8_12}
\showDOI{\tempurl}
\newblock
\shownote{Journal Abbreviation: Lecture Notes in Computer Science Publication Title: Lecture Notes in Computer Science}.


\bibitem[Dam et~al\mbox{.}(2018)]%
        {ExplainableSoftwareAnalytics}
\bibfield{author}{\bibinfo{person}{Hoa~Khanh Dam}, \bibinfo{person}{Truyen Tran}, {and} \bibinfo{person}{Aditya Ghose}.} \bibinfo{year}{2018}\natexlab{}.
\newblock \showarticletitle{Explainable software analytics}. In \bibinfo{booktitle}{\emph{Proceedings of the 40th {International} {Conference} on {Software} {Engineering}: {New} {Ideas} and {Emerging} {Results}}}. \bibinfo{publisher}{ACM}, \bibinfo{address}{Gothenburg Sweden}, \bibinfo{pages}{53--56}.
\newblock
\showISBNx{978-1-4503-5662-6}
\urldef\tempurl%
\url{https://doi.org/10.1145/3183399.3183424}
\showDOI{\tempurl}


\bibitem[Dietvorst et~al\mbox{.}(2015)]%
        {dietvorst:2015algorithm}
\bibfield{author}{\bibinfo{person}{Berkeley~J Dietvorst}, \bibinfo{person}{Joseph~P Simmons}, {and} \bibinfo{person}{Cade Massey}.} \bibinfo{year}{2015}\natexlab{}.
\newblock \showarticletitle{Algorithm aversion: people erroneously avoid algorithms after seeing them err.}
\newblock \bibinfo{journal}{\emph{Journal of Experimental Psychology: General}} \bibinfo{volume}{144}, \bibinfo{number}{1} (\bibinfo{year}{2015}), \bibinfo{pages}{114}.
\newblock


\bibitem[Dougherty({[n.\,d.]})]%
        {PartneringPeopleDeepLearning}
\bibfield{author}{\bibinfo{person}{Sean Dougherty}.} \bibinfo{year}{[n.\,d.]}\natexlab{}.
\newblock \showarticletitle{Partnering {People} with {Deep} {Learning} {Systems}: {Human} {Cognitive} {Effects} of {Explanations}}.
\newblock  (\bibinfo{year}{[n.\,d.]}), \bibinfo{pages}{231}.
\newblock


\bibitem[Ehsan et~al\mbox{.}(2023)]%
        {ehsan2023charting}
\bibfield{author}{\bibinfo{person}{Upol Ehsan}, \bibinfo{person}{Koustuv Saha}, \bibinfo{person}{Munmun De~Choudhury}, {and} \bibinfo{person}{Mark~O Riedl}.} \bibinfo{year}{2023}\natexlab{}.
\newblock \showarticletitle{Charting the Sociotechnical Gap in Explainable AI: A Framework to Address the Gap in XAI}.
\newblock \bibinfo{journal}{\emph{Proceedings of the ACM on Human-Computer Interaction}} \bibinfo{volume}{7}, \bibinfo{number}{CSCW1} (\bibinfo{year}{2023}), \bibinfo{pages}{1--32}.
\newblock


\bibitem[El-Sappagh and Elmogy(2015)]%
        {el2015case}
\bibfield{author}{\bibinfo{person}{Shaker~H El-Sappagh} {and} \bibinfo{person}{Mohammed Elmogy}.} \bibinfo{year}{2015}\natexlab{}.
\newblock \showarticletitle{Case based reasoning: Case representation methodologies}.
\newblock \bibinfo{journal}{\emph{International Journal of Advanced Computer Science and Applications}} \bibinfo{volume}{6}, \bibinfo{number}{11} (\bibinfo{year}{2015}), \bibinfo{pages}{192--208}.
\newblock


\bibitem[Fazio and Olson(2014)]%
        {fazio2014attitude}
\bibfield{author}{\bibinfo{person}{RH Fazio} {and} \bibinfo{person}{MA Olson}.} \bibinfo{year}{2014}\natexlab{}.
\newblock \showarticletitle{Attitude-behavior processes as a function of motivation and opportunity}.
\newblock \bibinfo{journal}{\emph{Dual Process Theories of the Social Mind. New York, NY: Guilford Press. Retrieved from http://www. unc. edu/\~{} dcameron/dualprocess. pdf}} (\bibinfo{year}{2014}).
\newblock


\bibitem[Feuillâtre et~al\mbox{.}(2020)]%
        {Similaritymeasuresattributeselectioncasebased}
\bibfield{author}{\bibinfo{person}{Hélène Feuillâtre}, \bibinfo{person}{Vincent Auffret}, \bibinfo{person}{Miguel Castro}, \bibinfo{person}{Florent Lalys}, \bibinfo{person}{Hervé Le~Breton}, \bibinfo{person}{Mireille Garreau}, {and} \bibinfo{person}{Pascal Haigron}.} \bibinfo{year}{2020}\natexlab{}.
\newblock \showarticletitle{Similarity measures and attribute selection for case-based reasoning in transcatheter aortic valve implantation}.
\newblock \bibinfo{journal}{\emph{PLOS ONE}} \bibinfo{volume}{15}, \bibinfo{number}{9} (\bibinfo{date}{Sept.} \bibinfo{year}{2020}), \bibinfo{pages}{e0238463}.
\newblock
\showISSN{1932-6203}
\urldef\tempurl%
\url{https://doi.org/10.1371/journal.pone.0238463}
\showDOI{\tempurl}


\bibitem[Galotti et~al\mbox{.}(2006)]%
        {DecisionmakingstylesreallifedecisionChoosing}
\bibfield{author}{\bibinfo{person}{Kathleen~M. Galotti}, \bibinfo{person}{Elizabeth Ciner}, \bibinfo{person}{Hope~E. Altenbaumer}, \bibinfo{person}{Heather~J. Geerts}, \bibinfo{person}{Allison Rupp}, {and} \bibinfo{person}{Julie Woulfe}.} \bibinfo{year}{2006}\natexlab{}.
\newblock \showarticletitle{Decision-making styles in a real-life decision: {Choosing} a college major}.
\newblock \bibinfo{journal}{\emph{Personality and Individual Differences}} \bibinfo{volume}{41}, \bibinfo{number}{4} (\bibinfo{date}{Sept.} \bibinfo{year}{2006}), \bibinfo{pages}{629--639}.
\newblock
\showISSN{01918869}
\urldef\tempurl%
\url{https://doi.org/10.1016/j.paid.2006.03.003}
\showDOI{\tempurl}


\bibitem[Grynaviski({[n.\,d.]})]%
        {Contrastscounterfactualscauses}
\bibfield{author}{\bibinfo{person}{Eric Grynaviski}.} \bibinfo{year}{[n.\,d.]}\natexlab{}.
\newblock \bibinfo{title}{Contrasts, counterfactuals, and causes}.
\newblock
\newblock


\bibitem[Guerra-Carrillo et~al\mbox{.}(2017)]%
        {Doeshighereducationhonecognitive}
\bibfield{author}{\bibinfo{person}{Belén Guerra-Carrillo}, \bibinfo{person}{Kiefer Katovich}, {and} \bibinfo{person}{Silvia~A. Bunge}.} \bibinfo{year}{2017}\natexlab{}.
\newblock \showarticletitle{Does higher education hone cognitive functioning and learning efficacy? {Findings} from a large and diverse sample}.
\newblock \bibinfo{journal}{\emph{PLOS ONE}} \bibinfo{volume}{12}, \bibinfo{number}{8} (\bibinfo{date}{Aug.} \bibinfo{year}{2017}), \bibinfo{pages}{e0182276}.
\newblock
\showISSN{1932-6203}
\urldef\tempurl%
\url{https://doi.org/10.1371/journal.pone.0182276}
\showDOI{\tempurl}


\bibitem[Harbers and Broekens({[n.\,d.]})]%
        {GuidelinesDevelopingExplainableCognitive}
\bibfield{author}{\bibinfo{person}{Maaike Harbers} {and} \bibinfo{person}{Joost Broekens}.} \bibinfo{year}{[n.\,d.]}\natexlab{}.
\newblock \showarticletitle{Guidelines for {Developing} {Explainable} {Cognitive} {Models}}.
\newblock  (\bibinfo{year}{[n.\,d.]}), \bibinfo{pages}{6}.
\newblock


\bibitem[Hilton(1990)]%
        {ConversationalProcessesCausalExplanation}
\bibfield{author}{\bibinfo{person}{Denis Hilton}.} \bibinfo{year}{1990}\natexlab{}.
\newblock \showarticletitle{Conversational processes and causal explanation}.
\newblock \bibinfo{journal}{\emph{Psychological Bulletin}}  \bibinfo{volume}{107} (\bibinfo{date}{Jan.} \bibinfo{year}{1990}), \bibinfo{pages}{65--81}.
\newblock
\urldef\tempurl%
\url{https://doi.org/10.1037/0033-2909.107.1.65}
\showDOI{\tempurl}


\bibitem[Hilton and John(2007)]%
        {hilton2007course}
\bibfield{author}{\bibinfo{person}{Denis~J Hilton} {and} \bibinfo{person}{L~McCLURE John}.} \bibinfo{year}{2007}\natexlab{}.
\newblock \showarticletitle{The course of events: counterfactuals, causal sequences, and explanation}.
\newblock In \bibinfo{booktitle}{\emph{The psychology of counterfactual thinking}}. \bibinfo{publisher}{Routledge}, \bibinfo{pages}{56--72}.
\newblock


\bibitem[Hilton and Slugoski(1986)]%
        {hilton1986knowledge}
\bibfield{author}{\bibinfo{person}{Denis~J Hilton} {and} \bibinfo{person}{Ben~R Slugoski}.} \bibinfo{year}{1986}\natexlab{}.
\newblock \showarticletitle{Knowledge-based causal attribution: The abnormal conditions focus model.}
\newblock \bibinfo{journal}{\emph{Psychological review}} \bibinfo{volume}{93}, \bibinfo{number}{1} (\bibinfo{year}{1986}), \bibinfo{pages}{75}.
\newblock


\bibitem[Hoffman et~al\mbox{.}(2019)]%
        {MetricsExplainableAIChallenges}
\bibfield{author}{\bibinfo{person}{Robert~R. Hoffman}, \bibinfo{person}{Shane~T. Mueller}, \bibinfo{person}{Gary Klein}, {and} \bibinfo{person}{Jordan Litman}.} \bibinfo{year}{2019}\natexlab{}.
\newblock \showarticletitle{Metrics for {Explainable} {AI}: {Challenges} and {Prospects}}.
\newblock \bibinfo{journal}{\emph{arXiv:1812.04608 [cs]}} (\bibinfo{date}{Feb.} \bibinfo{year}{2019}).
\newblock
\urldef\tempurl%
\url{http://arxiv.org/abs/1812.04608}
\showURL{%
\tempurl}
\newblock
\shownote{arXiv: 1812.04608}.


\bibitem[Jacobs et~al\mbox{.}(2021)]%
        {DesigningAITrustCollaborationTimeConstrained}
\bibfield{author}{\bibinfo{person}{Maia Jacobs}, \bibinfo{person}{Jeffrey He}, \bibinfo{person}{Melanie F.~Pradier}, \bibinfo{person}{Barbara Lam}, \bibinfo{person}{Andrew~C. Ahn}, \bibinfo{person}{Thomas~H. McCoy}, \bibinfo{person}{Roy~H. Perlis}, \bibinfo{person}{Finale Doshi-Velez}, {and} \bibinfo{person}{Krzysztof~Z. Gajos}.} \bibinfo{year}{2021}\natexlab{}.
\newblock \showarticletitle{Designing {AI} for {Trust} and {Collaboration} in {Time}-{Constrained} {Medical} {Decisions}: {A} {Sociotechnical} {Lens}}. In \bibinfo{booktitle}{\emph{Proceedings of the 2021 {CHI} {Conference} on {Human} {Factors} in {Computing} {Systems}}}. \bibinfo{publisher}{ACM}, \bibinfo{address}{Yokohama Japan}, \bibinfo{pages}{1--14}.
\newblock
\showISBNx{978-1-4503-8096-6}
\urldef\tempurl%
\url{https://doi.org/10.1145/3411764.3445385}
\showDOI{\tempurl}


\bibitem[Jepson and Ryan(2018)]%
        {jepson2018applying}
\bibfield{author}{\bibinfo{person}{Allan Jepson} {and} \bibinfo{person}{W~Gerard Ryan}.} \bibinfo{year}{2018}\natexlab{}.
\newblock \showarticletitle{Applying the motivation, opportunity, ability (MOA) model, and self-efficacy (SE) to better understand student engagement on undergraduate event management programs}.
\newblock \bibinfo{journal}{\emph{Event Management}} \bibinfo{volume}{22}, \bibinfo{number}{2} (\bibinfo{year}{2018}), \bibinfo{pages}{271--285}.
\newblock


\bibitem[Jim{\'e}nez-Luna et~al\mbox{.}(2020)]%
        {jimenez2020drug}
\bibfield{author}{\bibinfo{person}{Jos{\'e} Jim{\'e}nez-Luna}, \bibinfo{person}{Francesca Grisoni}, {and} \bibinfo{person}{Gisbert Schneider}.} \bibinfo{year}{2020}\natexlab{}.
\newblock \showarticletitle{Drug discovery with explainable artificial intelligence}.
\newblock \bibinfo{journal}{\emph{Nature Machine Intelligence}} \bibinfo{volume}{2}, \bibinfo{number}{10} (\bibinfo{year}{2020}), \bibinfo{pages}{573--584}.
\newblock


\bibitem[Jo et~al\mbox{.}(2023)]%
        {UnderstandingBenefitsChallengesDeployingConversationala}
\bibfield{author}{\bibinfo{person}{Eunkyung Jo}, \bibinfo{person}{Daniel~A. Epstein}, \bibinfo{person}{Hyunhoon Jung}, {and} \bibinfo{person}{Young-Ho Kim}.} \bibinfo{year}{2023}\natexlab{}.
\newblock \showarticletitle{Understanding the {Benefits} and {Challenges} of {Deploying} {Conversational} {AI} {Leveraging} {Large} {Language} {Models} for {Public} {Health} {Intervention}}. In \bibinfo{booktitle}{\emph{Proceedings of the 2023 {CHI} {Conference} on {Human} {Factors} in {Computing} {Systems}}}. \bibinfo{publisher}{ACM}, \bibinfo{address}{Hamburg Germany}, \bibinfo{pages}{1--16}.
\newblock
\showISBNx{978-1-4503-9421-5}
\urldef\tempurl%
\url{https://doi.org/10.1145/3544548.3581503}
\showDOI{\tempurl}


\bibitem[Kahneman(2011)]%
        {kahneman2011thinking}
\bibfield{author}{\bibinfo{person}{Daniel Kahneman}.} \bibinfo{year}{2011}\natexlab{}.
\newblock \bibinfo{booktitle}{\emph{Thinking, fast and slow}}.
\newblock \bibinfo{publisher}{macmillan}.
\newblock


\bibitem[Kim et~al\mbox{.}(2018)]%
        {TextualExplanationsSelfDrivingVehicles}
\bibfield{author}{\bibinfo{person}{Jinkyu Kim}, \bibinfo{person}{Anna Rohrbach}, \bibinfo{person}{Trevor Darrell}, \bibinfo{person}{John Canny}, {and} \bibinfo{person}{Zeynep Akata}.} \bibinfo{year}{2018}\natexlab{}.
\newblock \showarticletitle{Textual {Explanations} for {Self}-{Driving} {Vehicles}}.
\newblock In \bibinfo{booktitle}{\emph{Computer {Vision} – {ECCV} 2018}}, \bibfield{editor}{\bibinfo{person}{Vittorio Ferrari}, \bibinfo{person}{Martial Hebert}, \bibinfo{person}{Cristian Sminchisescu}, {and} \bibinfo{person}{Yair Weiss}} (Eds.). Vol.~\bibinfo{volume}{11206}. \bibinfo{publisher}{Springer International Publishing}, \bibinfo{address}{Cham}, \bibinfo{pages}{577--593}.
\newblock
\showISBNx{978-3-030-01215-1 978-3-030-01216-8}
\urldef\tempurl%
\url{https://doi.org/10.1007/978-3-030-01216-8_35}
\showDOI{\tempurl}
\newblock
\shownote{Series Title: Lecture Notes in Computer Science}.


\bibitem[Kizilcec(2016)]%
        {HowMuchInformationEffects}
\bibfield{author}{\bibinfo{person}{René~F. Kizilcec}.} \bibinfo{year}{2016}\natexlab{}.
\newblock \showarticletitle{How {Much} {Information}?: {Effects} of {Transparency} on {Trust} in an {Algorithmic} {Interface}}. In \bibinfo{booktitle}{\emph{Proceedings of the 2016 {CHI} {Conference} on {Human} {Factors} in {Computing} {Systems}}}. \bibinfo{publisher}{ACM}, \bibinfo{address}{San Jose California USA}, \bibinfo{pages}{2390--2395}.
\newblock
\showISBNx{978-1-4503-3362-7}
\urldef\tempurl%
\url{https://doi.org/10.1145/2858036.2858402}
\showDOI{\tempurl}


\bibitem[Kulesza et~al\mbox{.}(2013)]%
        {TooMuchTooLittlea}
\bibfield{author}{\bibinfo{person}{Todd Kulesza}, \bibinfo{person}{Simone Stumpf}, \bibinfo{person}{Margaret Burnett}, \bibinfo{person}{Sherry Yang}, \bibinfo{person}{Irwin Kwan}, {and} \bibinfo{person}{Weng-Keen Wong}.} \bibinfo{year}{2013}\natexlab{}.
\newblock \showarticletitle{Too much, too little, or just right? {Ways} explanations impact end users' mental models}. In \bibinfo{booktitle}{\emph{2013 {IEEE} {Symposium} on {Visual} {Languages} and {Human} {Centric} {Computing}}}. \bibinfo{publisher}{IEEE}, \bibinfo{address}{San Jose, CA, USA}, \bibinfo{pages}{3--10}.
\newblock
\showISBNx{978-1-4799-0369-6}
\urldef\tempurl%
\url{https://doi.org/10.1109/VLHCC.2013.6645235}
\showDOI{\tempurl}


\bibitem[Lai et~al\mbox{.}({[n.\,d.]})]%
        {ScienceHumanAIDecisionMakingSurvey}
\bibfield{author}{\bibinfo{person}{Vivian Lai}, \bibinfo{person}{Chacha Chen}, \bibinfo{person}{Q~Vera Liao}, \bibinfo{person}{Alison Smith-Renner}, {and} \bibinfo{person}{Chenhao Tan}.} \bibinfo{year}{[n.\,d.]}\natexlab{}.
\newblock \showarticletitle{Towards a {Science} of {Human}-{AI} {Decision} {Making}: {A} {Survey} of {Empirical} {Studies}}.
\newblock  (\bibinfo{year}{[n.\,d.]}), \bibinfo{pages}{36}.
\newblock


\bibitem[Lai et~al\mbox{.}(2023)]%
        {SelectiveExplanationsLeveragingHumanInput}
\bibfield{author}{\bibinfo{person}{Vivian Lai}, \bibinfo{person}{Yiming Zhang}, \bibinfo{person}{Chacha Chen}, \bibinfo{person}{Q.~Vera Liao}, {and} \bibinfo{person}{Chenhao Tan}.} \bibinfo{year}{2023}\natexlab{}.
\newblock \bibinfo{title}{Selective {Explanations}: {Leveraging} {Human} {Input} to {Align} {Explainable} {AI}}.
\newblock
\newblock
\urldef\tempurl%
\url{http://arxiv.org/abs/2301.09656}
\showURL{%
\tempurl}
\newblock
\shownote{arXiv:2301.09656 [cs]}.


\bibitem[Larasati et~al\mbox{.}(2020)]%
        {EffectExplanationStylesUser}
\bibfield{author}{\bibinfo{person}{Retno Larasati}, \bibinfo{person}{Anna~De Liddo}, {and} \bibinfo{person}{Enrico Motta}.} \bibinfo{year}{2020}\natexlab{}.
\newblock \showarticletitle{The {Effect} of {Explanation} {Styles} on {User}’s {Trust}}.
\newblock  (\bibinfo{year}{2020}), \bibinfo{pages}{7}.
\newblock


\bibitem[Li et~al\mbox{.}({[n.\,d.]})]%
        {HumanAIInteractionHealthcareThree}
\bibfield{author}{\bibinfo{person}{Yunzhi Li}, \bibinfo{person}{Liuping Wang}, \bibinfo{person}{Shuai Ma}, \bibinfo{person}{Xiangmin Fan}, \bibinfo{person}{Zijun Wang}, \bibinfo{person}{Junfeng Jiao}, {and} \bibinfo{person}{Dakuo Wang}.} \bibinfo{year}{[n.\,d.]}\natexlab{}.
\newblock \showarticletitle{Human-{AI} {Interaction} in {Healthcare}: {Three} {Case} {Studies} {About} {How} {Patient}(s) {And} {Doctors} {Interact} with {AI} in a {Multi}-{Tiers} {Healthcare} {Network}}.
\newblock  (\bibinfo{year}{[n.\,d.]}), \bibinfo{pages}{6}.
\newblock


\bibitem[Lipton(2018)]%
        {lipton2018mythos}
\bibfield{author}{\bibinfo{person}{Zachary~C Lipton}.} \bibinfo{year}{2018}\natexlab{}.
\newblock \showarticletitle{The mythos of model interpretability: In machine learning, the concept of interpretability is both important and slippery.}
\newblock \bibinfo{journal}{\emph{Queue}} \bibinfo{volume}{16}, \bibinfo{number}{3} (\bibinfo{year}{2018}), \bibinfo{pages}{31--57}.
\newblock


\bibitem[L{\"o}ckenhoff(2018)]%
        {lockenhoff2018aging}
\bibfield{author}{\bibinfo{person}{Corinna~E L{\"o}ckenhoff}.} \bibinfo{year}{2018}\natexlab{}.
\newblock \showarticletitle{Aging and decision-making: A conceptual framework for future research-a mini-review}.
\newblock \bibinfo{journal}{\emph{Gerontology}} \bibinfo{volume}{64}, \bibinfo{number}{2} (\bibinfo{year}{2018}), \bibinfo{pages}{140--148}.
\newblock


\bibitem[Logg et~al\mbox{.}(2019)]%
        {logg:2019algorithm}
\bibfield{author}{\bibinfo{person}{Jennifer~M Logg}, \bibinfo{person}{Julia~A Minson}, {and} \bibinfo{person}{Don~A Moore}.} \bibinfo{year}{2019}\natexlab{}.
\newblock \showarticletitle{Algorithm appreciation: People prefer algorithmic to human judgment}.
\newblock \bibinfo{journal}{\emph{Organizational Behavior and Human Decision Processes}}  \bibinfo{volume}{151} (\bibinfo{year}{2019}), \bibinfo{pages}{90--103}.
\newblock


\bibitem[MacInnis et~al\mbox{.}(1991)]%
        {EnhancingMeasuringConsumersMotivationOpportunity}
\bibfield{author}{\bibinfo{person}{Deborah~J. MacInnis}, \bibinfo{person}{Christine Moorman}, {and} \bibinfo{person}{Bernard~J. Jaworski}.} \bibinfo{year}{1991}\natexlab{}.
\newblock \showarticletitle{Enhancing and {Measuring} {Consumers}' {Motivation}, {Opportunity}, and {Ability} to {Process} {Brand} {Information} from {Ads}}.
\newblock \bibinfo{journal}{\emph{Journal of Marketing}} \bibinfo{volume}{55}, \bibinfo{number}{4} (\bibinfo{date}{Oct.} \bibinfo{year}{1991}), \bibinfo{pages}{32}.
\newblock
\showISSN{00222429}
\urldef\tempurl%
\url{https://doi.org/10.2307/1251955}
\showDOI{\tempurl}


\bibitem[Meguerdichian et~al\mbox{.}(2016)]%
        {Workingmemorylimitedimprovingknowledge}
\bibfield{author}{\bibinfo{person}{Michael Meguerdichian}, \bibinfo{person}{Katie Walker}, {and} \bibinfo{person}{Komal Bajaj}.} \bibinfo{year}{2016}\natexlab{}.
\newblock \showarticletitle{Working memory is limited: improving knowledge transfer by optimising simulation through cognitive load theory}.
\newblock \bibinfo{journal}{\emph{BMJ Simulation and Technology Enhanced Learning}} \bibinfo{volume}{2}, \bibinfo{number}{4} (\bibinfo{date}{Nov.} \bibinfo{year}{2016}), \bibinfo{pages}{131--138}.
\newblock
\showISSN{2056-6697}
\urldef\tempurl%
\url{https://doi.org/10.1136/bmjstel-2015-000098}
\showDOI{\tempurl}


\bibitem[Millecamp et~al\mbox{.}(2020)]%
        {CogitoergoquidEffectCognitive}
\bibfield{author}{\bibinfo{person}{Martijn Millecamp}, \bibinfo{person}{Robin Haveneers}, {and} \bibinfo{person}{Katrien Verbert}.} \bibinfo{year}{2020}\natexlab{}.
\newblock \showarticletitle{Cogito ergo quid? {The} {Effect} of {Cognitive} {Style} in a {Transparent} {Mobile} {Music} {Recommender} {System}}. In \bibinfo{booktitle}{\emph{Proceedings of the 28th {ACM} {Conference} on {User} {Modeling}, {Adaptation} and {Personalization}}}. \bibinfo{publisher}{ACM}, \bibinfo{address}{Genoa Italy}, \bibinfo{pages}{323--327}.
\newblock
\showISBNx{978-1-4503-6861-2}
\urldef\tempurl%
\url{https://doi.org/10.1145/3340631.3394871}
\showDOI{\tempurl}


\bibitem[Miller(2019)]%
        {miller2019explanation}
\bibfield{author}{\bibinfo{person}{Tim Miller}.} \bibinfo{year}{2019}\natexlab{}.
\newblock \showarticletitle{Explanation in artificial intelligence: Insights from the social sciences}.
\newblock \bibinfo{journal}{\emph{Artificial intelligence}}  \bibinfo{volume}{267} (\bibinfo{year}{2019}), \bibinfo{pages}{1--38}.
\newblock


\bibitem[Miller(2023a)]%
        {ExplainableAIDeadLongLive}
\bibfield{author}{\bibinfo{person}{Tim Miller}.} \bibinfo{year}{2023}\natexlab{a}.
\newblock \bibinfo{title}{Explainable {AI} is {Dead}, {Long} {Live} {Explainable} {AI}! {Hypothesis}-driven decision support}.
\newblock
\newblock
\urldef\tempurl%
\url{http://arxiv.org/abs/2302.12389}
\showURL{%
\tempurl}
\newblock
\shownote{arXiv:2302.12389 [cs]}.


\bibitem[Miller(2023b)]%
        {miller2023explainable}
\bibfield{author}{\bibinfo{person}{Tim Miller}.} \bibinfo{year}{2023}\natexlab{b}.
\newblock \showarticletitle{Explainable AI is Dead, Long Live Explainable AI! Hypothesis-driven Decision Support using Evaluative AI}. In \bibinfo{booktitle}{\emph{Proceedings of the 2023 ACM Conference on Fairness, Accountability, and Transparency}}. \bibinfo{pages}{333--342}.
\newblock


\bibitem[Nugent et~al\mbox{.}(2009)]%
        {GainingInsightCasebasedExplanation}
\bibfield{author}{\bibinfo{person}{Conor Nugent}, \bibinfo{person}{Dónal Doyle}, {and} \bibinfo{person}{Pádraig Cunningham}.} \bibinfo{year}{2009}\natexlab{}.
\newblock \showarticletitle{Gaining insight through case-based explanation}.
\newblock \bibinfo{journal}{\emph{Journal of Intelligent Information Systems}} \bibinfo{volume}{32}, \bibinfo{number}{3} (\bibinfo{date}{June} \bibinfo{year}{2009}), \bibinfo{pages}{267--295}.
\newblock
\showISSN{0925-9902, 1573-7675}
\urldef\tempurl%
\url{https://doi.org/10.1007/s10844-008-0069-0}
\showDOI{\tempurl}


\bibitem[Orru and Longo(2019)]%
        {EvolutionCognitiveLoadTheoryMeasurement}
\bibfield{author}{\bibinfo{person}{Giuliano Orru} {and} \bibinfo{person}{Luca Longo}.} \bibinfo{year}{2019}\natexlab{}.
\newblock \bibinfo{booktitle}{\emph{The {Evolution} of {Cognitive} {Load} {Theory} and the {Measurement} of {Its} {Intrinsic}, {Extraneous} and {Germane} {Loads}: {A} {Review}}}.
\newblock
\showISBNx{978-3-030-14272-8}
\urldef\tempurl%
\url{https://doi.org/10.1007/978-3-030-14273-5_3}
\showDOI{\tempurl}
\newblock
\shownote{Pages: 48}.


\bibitem[Pachur and Spaar(2015)]%
        {Domainspecificpreferencesintuitiondeliberationdecision}
\bibfield{author}{\bibinfo{person}{Thorsten Pachur} {and} \bibinfo{person}{Melanie Spaar}.} \bibinfo{year}{2015}\natexlab{}.
\newblock \showarticletitle{Domain-specific preferences for intuition and deliberation in decision making.}
\newblock \bibinfo{journal}{\emph{Journal of Applied Research in Memory and Cognition}} \bibinfo{volume}{4}, \bibinfo{number}{3} (\bibinfo{date}{Sept.} \bibinfo{year}{2015}), \bibinfo{pages}{303--311}.
\newblock
\showISSN{2211-369X, 2211-3681}
\urldef\tempurl%
\url{https://doi.org/10.1016/j.jarmac.2015.07.006}
\showDOI{\tempurl}


\bibitem[Pearl(2014)]%
        {DeductiveApproachCausalInference}
\bibfield{author}{\bibinfo{person}{Judea Pearl}.} \bibinfo{year}{2014}\natexlab{}.
\newblock \showarticletitle{The {Deductive} {Approach} to {Causal} {Inference}}.
\newblock \bibinfo{journal}{\emph{Journal of Causal Inference}} \bibinfo{volume}{2}, \bibinfo{number}{2} (\bibinfo{date}{Jan.} \bibinfo{year}{2014}).
\newblock
\showISSN{2193-3677, 2193-3685}
\urldef\tempurl%
\url{https://doi.org/10.1515/jci-2014-0016}
\showDOI{\tempurl}


\bibitem[Pennington(2012)]%
        {SocialCognition}
\bibfield{author}{\bibinfo{person}{Donald~C Pennington}.} \bibinfo{year}{2012}\natexlab{}.
\newblock \showarticletitle{Social {Cognition}}.
\newblock \bibinfo{journal}{\emph{Routledge}} (\bibinfo{year}{2012}).
\newblock


\bibitem[Qin et~al\mbox{.}(2019)]%
        {CounterfactualStoryReasoningGeneration}
\bibfield{author}{\bibinfo{person}{Lianhui Qin}, \bibinfo{person}{Antoine Bosselut}, \bibinfo{person}{Ari Holtzman}, \bibinfo{person}{Chandra Bhagavatula}, \bibinfo{person}{Elizabeth Clark}, {and} \bibinfo{person}{Yejin Choi}.} \bibinfo{year}{2019}\natexlab{}.
\newblock \bibinfo{title}{Counterfactual {Story} {Reasoning} and {Generation}}.
\newblock
\newblock
\urldef\tempurl%
\url{http://arxiv.org/abs/1909.04076}
\showURL{%
\tempurl}
\newblock
\shownote{arXiv:1909.04076 [cs]}.


\bibitem[Riefle et~al\mbox{.}(2022)]%
        {InfluenceCognitiveStylesUsersUnderstandinga}
\bibfield{author}{\bibinfo{person}{Lara Riefle}, \bibinfo{person}{Patrick Hemmer}, \bibinfo{person}{Carina Benz}, \bibinfo{person}{Michael Vössing}, {and} \bibinfo{person}{Jannik Pries}.} \bibinfo{year}{2022}\natexlab{}.
\newblock \bibinfo{title}{On the {Influence} of {Cognitive} {Styles} on {Users}' {Understanding} of {Explanations}}.
\newblock
\newblock
\urldef\tempurl%
\url{http://arxiv.org/abs/2210.02123}
\showURL{%
\tempurl}
\newblock
\shownote{arXiv:2210.02123 [cs]}.


\bibitem[Robertson et~al\mbox{.}(2021)]%
        {WaitWhyAssessingBehaviorExplanation}
\bibfield{author}{\bibinfo{person}{Justus Robertson}, \bibinfo{person}{Athanasios~Vasileios Kokkinakis}, \bibinfo{person}{Jonathan Hook}, \bibinfo{person}{Ben Kirman}, \bibinfo{person}{Florian Block}, \bibinfo{person}{Marian~F Ursu}, \bibinfo{person}{Sagarika Patra}, \bibinfo{person}{Simon Demediuk}, \bibinfo{person}{Anders Drachen}, {and} \bibinfo{person}{Oluseyi Olarewaju}.} \bibinfo{year}{2021}\natexlab{}.
\newblock \showarticletitle{Wait, {But} {Why}?: {Assessing} {Behavior} {Explanation} {Strategies} for {Real}-{Time} {Strategy} {Games}}. In \bibinfo{booktitle}{\emph{26th {International} {Conference} on {Intelligent} {User} {Interfaces}}}. \bibinfo{publisher}{ACM}, \bibinfo{address}{College Station TX USA}, \bibinfo{pages}{32--42}.
\newblock
\showISBNx{978-1-4503-8017-1}
\urldef\tempurl%
\url{https://doi.org/10.1145/3397481.3450699}
\showDOI{\tempurl}


\bibitem[Roth-Berghofer(2004)]%
        {ExplanationsCaseBasedReasoningFoundational}
\bibfield{author}{\bibinfo{person}{Thomas Roth-Berghofer}.} \bibinfo{year}{2004}\natexlab{}.
\newblock \bibinfo{booktitle}{\emph{Explanations and {Case}-{Based} {Reasoning}: {Foundational} {Issues}}}. Vol.~\bibinfo{volume}{3155}.
\newblock
\showISBNx{978-3-540-22882-0}
\urldef\tempurl%
\url{https://doi.org/10.1007/978-3-540-28631-8_29}
\showDOI{\tempurl}
\newblock
\shownote{Pages: 403}.


\bibitem[Schraagen et~al\mbox{.}(2020)]%
        {TrustingXAIEffectsdifferenttypes}
\bibfield{author}{\bibinfo{person}{Jan~Maarten Schraagen}, \bibinfo{person}{Pia Elsasser}, \bibinfo{person}{Hanna Fricke}, \bibinfo{person}{Marleen Hof}, {and} \bibinfo{person}{Fabyen Ragalmuto}.} \bibinfo{year}{2020}\natexlab{}.
\newblock \showarticletitle{Trusting the {X} in {XAI}: {Effects} of different types of explanations by a self-driving car on trust, explanation satisfaction and mental models}.
\newblock \bibinfo{journal}{\emph{Proceedings of the Human Factors and Ergonomics Society Annual Meeting}} \bibinfo{volume}{64}, \bibinfo{number}{1} (\bibinfo{date}{Dec.} \bibinfo{year}{2020}), \bibinfo{pages}{339--343}.
\newblock
\showISSN{2169-5067, 1071-1813}
\urldef\tempurl%
\url{https://doi.org/10.1177/1071181320641077}
\showDOI{\tempurl}


\bibitem[Scott and Bruce(1995)]%
        {DecisionMakingStyleDevelopmentAssessmentNew}
\bibfield{author}{\bibinfo{person}{Susanne~G. Scott} {and} \bibinfo{person}{Reginald~A. Bruce}.} \bibinfo{year}{1995}\natexlab{}.
\newblock \showarticletitle{Decision-{Making} {Style}: {The} {Development} and {Assessment} of a {New} {Measure}}.
\newblock \bibinfo{journal}{\emph{Educational and Psychological Measurement}} \bibinfo{volume}{55}, \bibinfo{number}{5} (\bibinfo{date}{Oct.} \bibinfo{year}{1995}), \bibinfo{pages}{818--831}.
\newblock
\showISSN{0013-1644, 1552-3888}
\urldef\tempurl%
\url{https://doi.org/10.1177/0013164495055005017}
\showDOI{\tempurl}


\bibitem[Sendak et~al\mbox{.}(2020)]%
        {humanbodyblackboxsupporting}
\bibfield{author}{\bibinfo{person}{Mark Sendak}, \bibinfo{person}{Madeleine~Clare Elish}, \bibinfo{person}{Michael Gao}, \bibinfo{person}{Joseph Futoma}, \bibinfo{person}{William Ratliff}, \bibinfo{person}{Marshall Nichols}, \bibinfo{person}{Armando Bedoya}, \bibinfo{person}{Suresh Balu}, {and} \bibinfo{person}{Cara O'Brien}.} \bibinfo{year}{2020}\natexlab{}.
\newblock \showarticletitle{"{The} human body is a black box": supporting clinical decision-making with deep learning}. In \bibinfo{booktitle}{\emph{Proceedings of the 2020 {Conference} on {Fairness}, {Accountability}, and {Transparency}}}. \bibinfo{publisher}{ACM}, \bibinfo{address}{Barcelona Spain}, \bibinfo{pages}{99--109}.
\newblock
\showISBNx{978-1-4503-6936-7}
\urldef\tempurl%
\url{https://doi.org/10.1145/3351095.3372827}
\showDOI{\tempurl}


\bibitem[Sheh({[n.\,d.]})]%
        {DifferentXAIDifferentHRI}
\bibfield{author}{\bibinfo{person}{Raymond~K Sheh}.} \bibinfo{year}{[n.\,d.]}\natexlab{}.
\newblock \showarticletitle{Different {XAI} for {Different} {HRI}}.
\newblock  (\bibinfo{year}{[n.\,d.]}), \bibinfo{pages}{4}.
\newblock


\bibitem[Spicer and Sadler‐Smith(2005)]%
        {examinationgeneraldecisionmakingstyle}
\bibfield{author}{\bibinfo{person}{David~P. Spicer} {and} \bibinfo{person}{Eugene Sadler‐Smith}.} \bibinfo{year}{2005}\natexlab{}.
\newblock \showarticletitle{An examination of the general decision making style questionnaire in two {UK} samples}.
\newblock \bibinfo{journal}{\emph{Journal of Managerial Psychology}} \bibinfo{volume}{20}, \bibinfo{number}{2} (\bibinfo{date}{March} \bibinfo{year}{2005}), \bibinfo{pages}{137--149}.
\newblock
\showISSN{0268-3946}
\urldef\tempurl%
\url{https://doi.org/10.1108/02683940510579777}
\showDOI{\tempurl}


\bibitem[Sridharan and Meadows(2019)]%
        {TheoryExplanationsHumanRobotCollaboration}
\bibfield{author}{\bibinfo{person}{Mohan Sridharan} {and} \bibinfo{person}{Ben Meadows}.} \bibinfo{year}{2019}\natexlab{}.
\newblock \showarticletitle{Towards a {Theory} of {Explanations} for {Human}–{Robot} {Collaboration}}.
\newblock \bibinfo{journal}{\emph{KI - Künstliche Intelligenz}} \bibinfo{volume}{33}, \bibinfo{number}{4} (\bibinfo{date}{Dec.} \bibinfo{year}{2019}), \bibinfo{pages}{331--342}.
\newblock
\showISSN{0933-1875, 1610-1987}
\urldef\tempurl%
\url{https://doi.org/10.1007/s13218-019-00616-y}
\showDOI{\tempurl}


\bibitem[Stange et~al\mbox{.}(2019)]%
        {SelfExplainingSocialRobotsVerbal}
\bibfield{author}{\bibinfo{person}{Sonja Stange}, \bibinfo{person}{Hendrik Buschmeier}, \bibinfo{person}{Teena Hassan}, \bibinfo{person}{Christopher Ritter}, {and} \bibinfo{person}{Stefan Kopp}.} \bibinfo{year}{2019}\natexlab{}.
\newblock \showarticletitle{Towards {Self}-{Explaining} {Social} {Robots}: {Verbal} {Explanation} {Strategies} for a {Needs}-{Based} {Architecture}}.
\newblock  (\bibinfo{year}{2019}), \bibinfo{pages}{6}.
\newblock


\bibitem[Stolpmann and Wess(1999)]%
        {stolpmann1999optimierung}
\bibfield{author}{\bibinfo{person}{Markus Stolpmann} {and} \bibinfo{person}{Stefan Wess}.} \bibinfo{year}{1999}\natexlab{}.
\newblock \bibinfo{booktitle}{\emph{Optimierung der Kundenbeziehung mit CBR-Systemen: Intelligente Systeme f{\"u}r E-Commerce und Support}}.
\newblock \bibinfo{publisher}{Addison-Wesley-Longman}.
\newblock


\bibitem[Sun et~al\mbox{.}(2021)]%
        {CapturingTrendsApplicationsIssues}
\bibfield{author}{\bibinfo{person}{Lingyun Sun}, \bibinfo{person}{Zhuoshu Li}, \bibinfo{person}{Yuyang Zhang}, \bibinfo{person}{Yanzhen Liu}, \bibinfo{person}{Shanghua Lou}, {and} \bibinfo{person}{Zhibin Zhou}.} \bibinfo{year}{2021}\natexlab{}.
\newblock \showarticletitle{Capturing the {Trends}, {Applications}, {Issues}, and {Potential} {Strategies} of {Designing} {Transparent} {AI} {Agents}}. In \bibinfo{booktitle}{\emph{Extended {Abstracts} of the 2021 {CHI} {Conference} on {Human} {Factors} in {Computing} {Systems}}}. \bibinfo{publisher}{ACM}, \bibinfo{address}{Yokohama Japan}, \bibinfo{pages}{1--8}.
\newblock
\showISBNx{978-1-4503-8095-9}
\urldef\tempurl%
\url{https://doi.org/10.1145/3411763.3451819}
\showDOI{\tempurl}


\bibitem[Sweller({[n.\,d.]})]%
        {CognitiveArchitectureInstructionalDesign}
\bibfield{author}{\bibinfo{person}{John Sweller}.} \bibinfo{year}{[n.\,d.]}\natexlab{}.
\newblock \showarticletitle{Cognitive {Architecture} and {Instructional} {Design}}.
\newblock  (\bibinfo{year}{[n.\,d.]}).
\newblock


\bibitem[Tsai et~al\mbox{.}(2021)]%
        {ExploringPromotingDiagnosticTransparency}
\bibfield{author}{\bibinfo{person}{Chun-Hua Tsai}, \bibinfo{person}{Yue You}, \bibinfo{person}{Xinning Gui}, \bibinfo{person}{Yubo Kou}, {and} \bibinfo{person}{John~M Carroll}.} \bibinfo{year}{2021}\natexlab{}.
\newblock \showarticletitle{Exploring and {Promoting} {Diagnostic} {Transparency} and {Explainability} in {Online} {Symptom} {Checkers}}.
\newblock  (\bibinfo{year}{2021}), \bibinfo{pages}{18}.
\newblock


\bibitem[Van~Bouwel and Weber(2002)]%
        {RemoteCausesBadExplanations}
\bibfield{author}{\bibinfo{person}{Jeroen Van~Bouwel} {and} \bibinfo{person}{Erik Weber}.} \bibinfo{year}{2002}\natexlab{}.
\newblock \showarticletitle{Remote {Causes}, {Bad} {Explanations}?}
\newblock \bibinfo{journal}{\emph{Journal for the Theory of Social Behaviour}} \bibinfo{volume}{32}, \bibinfo{number}{4} (\bibinfo{date}{Dec.} \bibinfo{year}{2002}), \bibinfo{pages}{437--449}.
\newblock
\showISSN{0021-8308, 1468-5914}
\urldef\tempurl%
\url{https://doi.org/10.1111/1468-5914.00197}
\showDOI{\tempurl}


\bibitem[Vasconcelos et~al\mbox{.}(2023)]%
        {ExplanationsCanReduceOverrelianceAI}
\bibfield{author}{\bibinfo{person}{Helena Vasconcelos}, \bibinfo{person}{Matthew Jörke}, \bibinfo{person}{Madeleine Grunde-McLaughlin}, \bibinfo{person}{Tobias Gerstenberg}, \bibinfo{person}{Michael~S. Bernstein}, {and} \bibinfo{person}{Ranjay Krishna}.} \bibinfo{year}{2023}\natexlab{}.
\newblock \showarticletitle{Explanations {Can} {Reduce} {Overreliance} on {AI} {Systems} {During} {Decision}-{Making}}.
\newblock \bibinfo{journal}{\emph{Proceedings of the ACM on Human-Computer Interaction}} \bibinfo{volume}{7}, \bibinfo{number}{CSCW1} (\bibinfo{date}{April} \bibinfo{year}{2023}), \bibinfo{pages}{1--38}.
\newblock
\showISSN{2573-0142}
\urldef\tempurl%
\url{https://doi.org/10.1145/3579605}
\showDOI{\tempurl}


\bibitem[Vereschak et~al\mbox{.}(2021)]%
        {HowEvaluateTrustAIAssistedDecision}
\bibfield{author}{\bibinfo{person}{Oleksandra Vereschak}, \bibinfo{person}{Gilles Bailly}, {and} \bibinfo{person}{Baptiste Caramiaux}.} \bibinfo{year}{2021}\natexlab{}.
\newblock \showarticletitle{How to {Evaluate} {Trust} in {AI}-{Assisted} {Decision} {Making}? {A} {Survey} of {Empirical} {Methodologies}}.
\newblock \bibinfo{journal}{\emph{Proceedings of the ACM on Human-Computer Interaction}} \bibinfo{volume}{5}, \bibinfo{number}{CSCW2} (\bibinfo{date}{Oct.} \bibinfo{year}{2021}), \bibinfo{pages}{1--39}.
\newblock
\showISSN{2573-0142}
\urldef\tempurl%
\url{https://doi.org/10.1145/3476068}
\showDOI{\tempurl}


\bibitem[Vilone and Longo(2020)]%
        {vilone2020explainable}
\bibfield{author}{\bibinfo{person}{Giulia Vilone} {and} \bibinfo{person}{Luca Longo}.} \bibinfo{year}{2020}\natexlab{}.
\newblock \showarticletitle{Explainable artificial intelligence: a systematic review}.
\newblock \bibinfo{journal}{\emph{arXiv preprint arXiv:2006.00093}} (\bibinfo{year}{2020}).
\newblock


\bibitem[Wachter et~al\mbox{.}(2017)]%
        {CounterfactualExplanationsOpeningBlack}
\bibfield{author}{\bibinfo{person}{Sandra Wachter}, \bibinfo{person}{Brent Mittelstadt}, {and} \bibinfo{person}{Chris Russell}.} \bibinfo{year}{2017}\natexlab{}.
\newblock \showarticletitle{Counterfactual {Explanations} {Without} {Opening} the {Black} {Box}: {Automated} {Decisions} and the {GDPR}}.
\newblock \bibinfo{journal}{\emph{SSRN Electronic Journal}} (\bibinfo{year}{2017}).
\newblock
\showISSN{1556-5068}
\urldef\tempurl%
\url{https://doi.org/10.2139/ssrn.3063289}
\showDOI{\tempurl}


\bibitem[Wang and Yin(2021)]%
        {AreExplanationsHelpfulComparative}
\bibfield{author}{\bibinfo{person}{Xinru Wang} {and} \bibinfo{person}{Ming Yin}.} \bibinfo{year}{2021}\natexlab{}.
\newblock \showarticletitle{Are {Explanations} {Helpful}? {A} {Comparative} {Study} of the {Effects} of {Explanations} in {AI}-{Assisted} {Decision}-{Making}}.
\newblock  (\bibinfo{year}{2021}), \bibinfo{pages}{11}.
\newblock


\bibitem[Watson(1999)]%
        {SurveyCBRApplicationAreas}
\bibfield{author}{\bibinfo{person}{Ian Watson}.} \bibinfo{year}{1999}\natexlab{}.
\newblock \showarticletitle{Survey of {CBR} {Application} {Areas}}.
\newblock  (\bibinfo{year}{1999}).
\newblock


\bibitem[Weiner(1972)]%
        {weiner1972attribution}
\bibfield{author}{\bibinfo{person}{Bernard Weiner}.} \bibinfo{year}{1972}\natexlab{}.
\newblock \showarticletitle{Attribution theory, achievement motivation, and the educational process}.
\newblock \bibinfo{journal}{\emph{Review of educational research}} \bibinfo{volume}{42}, \bibinfo{number}{2} (\bibinfo{year}{1972}), \bibinfo{pages}{203--215}.
\newblock


\bibitem[Woodward({[n.\,d.]})]%
        {PsychologicalStudiesCausalCounterfactualReasoning}
\bibfield{author}{\bibinfo{person}{James Woodward}.} \bibinfo{year}{[n.\,d.]}\natexlab{}.
\newblock \showarticletitle{1 1 {Psychological} {Studies} of {Causal} and {Counterfactual} {Reasoning} }.
\newblock  (\bibinfo{year}{[n.\,d.]}).
\newblock


\bibitem[Woodward(2006)]%
        {woodward2006sensitive}
\bibfield{author}{\bibinfo{person}{James Woodward}.} \bibinfo{year}{2006}\natexlab{}.
\newblock \showarticletitle{Sensitive and insensitive causation}.
\newblock \bibinfo{journal}{\emph{The Philosophical Review}} \bibinfo{volume}{115}, \bibinfo{number}{1} (\bibinfo{year}{2006}), \bibinfo{pages}{1--50}.
\newblock


\bibitem[Xie et~al\mbox{.}(2020)]%
        {CheXplainEnablingPhysiciansExploreUnderstand}
\bibfield{author}{\bibinfo{person}{Yao Xie}, \bibinfo{person}{Melody Chen}, \bibinfo{person}{David Kao}, \bibinfo{person}{Ge Gao}, {and} \bibinfo{person}{Xiang~'Anthony' Chen}.} \bibinfo{year}{2020}\natexlab{}.
\newblock \showarticletitle{{CheXplain}: {Enabling} {Physicians} to {Explore} and {Understand} {Data}-{Driven}, {AI}-{Enabled} {Medical} {Imaging} {Analysis}}. In \bibinfo{booktitle}{\emph{Proceedings of the 2020 {CHI} {Conference} on {Human} {Factors} in {Computing} {Systems}}}. \bibinfo{publisher}{ACM}, \bibinfo{address}{Honolulu HI USA}, \bibinfo{pages}{1--13}.
\newblock
\showISBNx{978-1-4503-6708-0}
\urldef\tempurl%
\url{https://doi.org/10.1145/3313831.3376807}
\showDOI{\tempurl}


\bibitem[Zhang and Lim(2022)]%
        {RelatableExplainableAIPerceptualProcess}
\bibfield{author}{\bibinfo{person}{Wencan Zhang} {and} \bibinfo{person}{Brian~Y Lim}.} \bibinfo{year}{2022}\natexlab{}.
\newblock \showarticletitle{Towards {Relatable} {Explainable} {AI} with the {Perceptual} {Process}}. In \bibinfo{booktitle}{\emph{{CHI} {Conference} on {Human} {Factors} in {Computing} {Systems}}}. \bibinfo{publisher}{ACM}, \bibinfo{address}{New Orleans LA USA}, \bibinfo{pages}{1--24}.
\newblock
\showISBNx{978-1-4503-9157-3}
\urldef\tempurl%
\url{https://doi.org/10.1145/3491102.3501826}
\showDOI{\tempurl}


\end{thebibliography}
